\documentclass[iop, apj, numberedappendix, twocolappendix, appendixfloats]{emulateapj} 
\usepackage{times}
\usepackage{ amssymb }

\usepackage{color}

\renewcommand{\b}{\hat{\bb{b}}}
\newcommand{\be}{\begin{eqnarray}}
\newcommand{\en}{\end{eqnarray}}
\newcommand{\pa}{\partial}
\newcommand{\f}{\frac}
\newcommand{\tci}{\tau^{-1}_{\rm c}} 
\newcommand{\tdi}{\tau^{-1}_{\rm d}}
\newcommand{\tvi}{\tau^{-1}_{\rm v}}

\newcommand{\K}{\mathcal{K}}
\newcommand{\od}{\omega_{\rm dyn}}

\newcommand\bb[1]{\mbox{\boldmath{$#1$}}}
 
\newcommand\bcdot{\bb{\cdot}}
\newcommand\btimes{\bb{\times}}

\begin{document}

\title{THE STABILITY OF WEAKLY COLLISIONAL PLASMAS WITH THERMAL AND COMPOSITION GRADIENTS}

\author{Martin~ E. Pessah$^1$ and Sagar Chakraborty$^{1,2}$}
\affil{$^1$Niels Bohr International Academy, Niels Bohr Institute, Blegdamsvej 17, DK-2100 Copenhagen \O, Denmark; 
\url{mpessah@nbi.dk} \\$^2$Department of Physics, Indian Institute of Technology, Kanpur, U.P.-208016, India; 
\url{sagarc@iitk.ac.in}}

\shorttitle{THE STABILITY OF WEAKLY COLLISIONAL PLASMAS}
\shortauthors{Pessah \& Chakraborty}

\begin{abstract}
Over the last decade, substantial efforts have been devoted to
understanding the stability properties, transport phenomena, and
long-term evolution of weakly collisional, magnetized plasmas which
are stratified in temperature.  The insights gained via these studies
have led to a significant improvement of our understanding of the
processes that determine the physical evolution and observational
properties of the intracluster medium (ICM) permeating galaxy
clusters.  These studies have been carried out under the assumption
that the ICM is a homogeneous medium. This, however, might not be a
good approximation if heavy elements are able to sediment in the inner
region of the galaxy cluster. Motivated by the need to obtain a more
complete picture of the dynamical properties of the ICM, we analyze
the stability of a weakly collisional, magnetized plane-parallel
atmosphere which is stratified in both temperature and composition.
This allows us to discuss for the first time the dynamics of
weakly collisional environments where heat conduction, momentum
transport, and ion-diffusion are anisotropic with respect to the
direction of the magnetic field. We show that, depending on the
relative signs and magnitudes of the gradients in the temperature and
the mean molecular weight, the plasma can be subject to a wide variety
of unstable modes which include modifications to the magnetothermal
instability (MTI), the heat-flux-driven buoyancy instability (HBI),
and overstable gravity modes previously studied in homogeneous media.
We also find that there are new modes which are driven by heat
conduction and particle diffusion.  We discuss the astrophysical
implications of our findings for a representative galaxy cluster where
helium has sedimented. Our findings suggest that the core insulation
that results from the magnetic field configurations that arise as a
natural consequence of the HBI, which would be MTI stable in a
homogeneous medium, could be alleviated if the mean molecular weight
gradient is steep enough, i.e., $(\nabla \mu)/\mu > (\nabla
T)/T$. This study constitutes a first step toward understanding the
interaction between magnetic turbulence and the diffusion of heavy
elements, and its consequences for the long-term evolution and
observational signatures of the ICM in galaxy clusters.
\end{abstract}


\section{Introduction}
\label{sec:introduction}

Despite the fact that magnetic fields in galaxy clusters are too weak
to be mechanically important, they can play a fundamental role in the
dynamical stability of the dilute gas by channeling the transport of
heat, momentum, and particles. The weakly collisional character of the hot 
intracluster medium (ICM), which is generically characterized by stable entropy 
gradients according to Schwarzschild's criterion \citep{2005A&A...433..101P,
  2009ApJS..182...12C}, enables the action of magnetic instabilities
that are sensitive to temperature gradients
\citep{2000ApJ...534..420B, 2004ApJ...616..857B}.  In particular, the
magneto-thermal instability (MTI) exhibits the fastest growing modes 
when magnetic field lines are orthogonal to a temperature gradient 
parallel to the gravitational field \citep{2001ApJ...562..909B}, whereas 
the heat-flux-driven buoyancy instability (HBI) does so when magnetic field 
lines are parallel to a temperature gradient which is anti-parallel to the 
gravitational field \citep{2008ApJ...673..758Q}. 

While the landscape of thermal instabilities that render homogeneous,
dilute plasmas unstable has been well explored
\citep{2011MNRAS.417..602K}, and even extended to account for the
effects of cosmic rays \citep{2006ApJ...642..140C, 2010ApJ...720..652S}, very little is
known about the effects that composition gradients can have on the
stability of the dilute ICM.  If magnetic fields do not prevent the
efficient diffusion of ions \citep{2001ApJ...562L.129N,
  2003MNRAS.342L...5C, 2004MNRAS.349L..13C} then the gradients in mean
molecular weight can be as important as the gradients in temperature
(see Section~\ref{sec:discussion} and \citealt{2000ApJ...529L...1Q,
  2009ApJ...693..839P, 2010MNRAS.401.1360S, 2011A&A...533A...6B}) and
provide another source of free energy to feed instabilities.  In order
to obtain a more complete picture of the stability properties of the
ICM, it is thus important to relax the assumption of a homogeneous
medium.

As a first step toward understanding the role of composition
gradients in the stability of dilute plasmas, such as the ICM, we
analyze the stability of a weakly magnetized plane-parallel atmosphere
where magnetic fields play a key role by channeling the conduction of
heat, transport of momentum, and the diffusion of ions.  
Our analysis generalizes previous studies on the MTI \citep{2001ApJ...562..909B}, 
the HBI  \citep{2008ApJ...673..758Q}, and overstable gravity modes 
\citep{2010ApJ...720L..97B}, and reveals the subtle roles played by the 
temperature and the composition gradients in determining the stability of the plasma.

The outline of the paper is as follows.
In Section \ref{sec:model} we describe the plasma model for a dilute binary mixture of ions.
In Section \ref{sec:stability} we perform the linear mode analysis
and we obtain the general dispersion relation that governs the linear dynamics
of a weakly magnetized medium which is stratified in temperature and composition.
We analyze in detail the stability of the plasma in the regimes where conduction
across a given scale is, respectively, fast and slow compared to the dynamical timescale 
in Sections \ref{sec:fast-conduction} and \ref{sec:slow-conduction}. 
In Section \ref{sec:physics} we describe the physics driving the most relevant instabilities.
We discuss the astrophysical implications of this study in Section  \ref{sec:discussion}.


\begin{figure}[t]
\begin{center}
  \includegraphics[width=0.4\textwidth,trim=0 0 0 0]{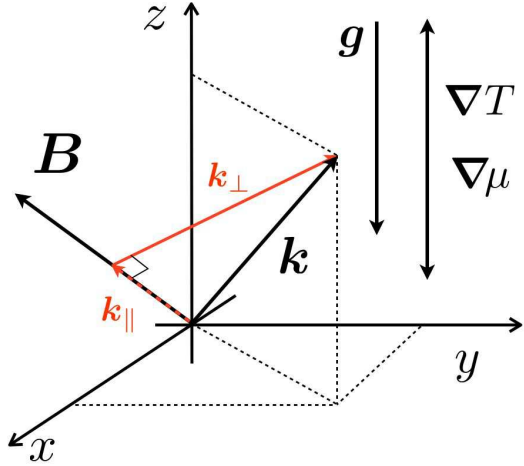}
  \caption{Schematic representation of the geometry involved in the
    stability analysis of the dilute, magnetized plane-parallel atmosphere with
    density, temperature, and composition gradients. The symbols
    $\parallel$ and $\bot$ label the directions parallel and
    perpendicular to the magnetic field, which is assumed to lie on
    the $x$--$z$ plane, without loss of generality.}
  \label{fig:geometry}
\end{center}
\end{figure}

\section{Model for the Multi-component, Dilute Atmosphere}
\label{sec:model}

\subsection{General Considerations for the Plasma Model}

In order to highlight the physical phenomena that emerge when
composition gradients are accounted for, we focus our attention on a
dilute binary mixture (e.g., hydrogen and helium)\footnote{
This analysis can be generalized to consider $N$ species.}
in a fixed gravitational field described by\footnote{For the sake of
  simplicity, we do not consider here the effects of
  thermo-diffusion, baro-diffusion, etc. \citep{1959flme.book.....L}.}
\be &&\f{\pa \rho}{\pa t}+\bb{\nabla}\bcdot(\rho \bb{v})=0 \,,
\label{eq:rho}\\
&&\f{\partial}{\partial t}(\rho \bb{v}) + \bb{\nabla} \bcdot \left(\rho \bb{v} \bb{v} + \mathsf{P} + 
\f{B^2}{8\pi}\mathsf{I} - \f{{B^2}}{4\pi}\hat{\bb{b}}\hat{\bb{b}}\right) = \rho\bb{g} \,, \,\,\,\,
\label{eq:v}\\
&&\f{\pa \bb{B}}{\pa t}=\bb{\nabla}\btimes(\bb{v}\btimes\bb{B}) \,,
\label{eq:b}\\
&& \f{P}{\gamma-1} \f{d}{dt}(\ln{P\rho^{-\gamma}})=
(p_\bot-p_\parallel)\f{d}{dt}\ln\f{B}{\rho^{2/3}} 
- \bb{\nabla} \bcdot \bb{Q}_{\rm s} 
 \,, \,\,\,\,\,\,\,
\label{eq:S} \\
&&\f{dc}{dt} =-\bb{\nabla}\bcdot\bb{Q}_{\rm c} \,.
\label{eq:c}
\en 
Here, the Lagrangian and Eulerian derivatives are related via
$d/dt \equiv \pa/\pa t + \bb{v} \bcdot \nabla$, $\rho$ is the mass
density, $\bb{v}$ is the fluid velocity, $\bb{g}$ is the gravitational
acceleration, $\gamma$ is the adiabatic index, and $\mathsf{I}$ stands
for the $3 \times 3$ identity matrix.  The symbols $\bot$ and
$\parallel$ refer respectively to the directions perpendicular and parallel to the
magnetic field $\bb{B}$, see Figure~\ref{fig:geometry}, whose
direction is given by the versor $\hat{\bb{b}}\equiv \bb{B}/B = (b_x,
0, b_z)$. 
The first term on the right-hand side of Equation (\ref{eq:S}) accounts for 
entropy production due to viscous heating in a weakly collisional 
magnetized plasma (see, e.g., \citealt{1985JGR....90.7620H}).
These equations have been considered in previous works 
investigating the dynamics of the weakly collisional ICM
with a single ion species, i.e., in the case where the concentration of the
other ion species is $c=0$, see, e.g., \citet{2011MNRAS.417..602K, 
2012MNRAS.422..704P, 2012ApJ...754..122K}, and references therein.

Equations~(\ref{eq:rho})--(\ref{eq:c}) describe the dynamics of a
dilute binary mixture in the low-collisionality regime and they differ from
standard MHD in three important respects.

(\emph{i}) In a weakly collisional magnetized plasma the pressure tensor 
$\mathsf{P} \equiv p_\bot \mathsf{I} + (p_\parallel - p_\bot) \hat{\bb{b}} 
\hat{\bb{b}}$ is anisotropic.
If the frequency of ion collisions $\nu_{ii}$ in single ion species 
magneto-fluid is large compared to the rate of change $d/dt$ of all 
the fields involved, then the anisotropic part of the pressure tensor 
is small compared to its isotropic part 
$P\equiv 2p_\bot/3 + p_\parallel/3$ and (see, e.g., 
\citealt{1985JGR....90.7620H, 2005ApJ...629..139S})
\be
\frac{|p_\parallel - p_\bot|}{P}
 = \frac{3}{\nu_{ii}} \, \left|\frac{d}{dt}(\ln B \rho^{-2/3})\right| \ll 1 \,.
\en
The anisotropic component of the pressure tensor in the momentum equation 
gives rise to the phenomenon known as Braginskii viscosity. For small pressure
anisotropy,\footnote{
In principle, appropriate closure approximations, see, e.g., \cite{1997PhPl....4.3974S}
may be adopted to address the high-collisionality regime starting from 
the Chew-Goldberger-Low, or CGL, limit \citep{1956RSPSA.236..112C}. 
A detailed application of this approach can be found in the context 
of accretion disks in \cite{2003ApJ...596.1121S}.} this contribution is usually written as 
\be p_\parallel
- p_\bot = 3 \eta_0 \left(\hat{\bb{b}} \hat{\bb{b}} - \frac{1}{3}
  \mathsf{I} \right) : \bb{\nabla} \bb{v} \,,
\label{eq:nu_braginskii}
\en
where $\eta_0$ is the largest of the coefficients in 
the viscous stress tensor derived by \citet{1965RvPP....1..205B}.
In order to account for the effects of collisions between ions of
different species in the binary mixture, we replace the $\nu_{ii}$
by an effective ion-ion collision frequency $\nu_{ii}^{\rm eff}$,
which we define in Appendix \ref{app:tdi-vs-tvi}.

(\emph{ii}) Heat flows mainly along magnetic field lines, because the
electron mean free path is large compared to its Larmor radius. This
process is modeled by the second term on the right-hand side of
Equation~(\ref{eq:S}) via 
\be
\bb{Q}_{\rm s} \equiv-\chi\b(\b\bcdot \bb{\nabla})T \,,
\en
where $T$ is the plasma temperature, assumed to be the
same for ions and electrons, and $\chi$ is the thermal 
conductivity predominately due to electrons 
\citep{1962pfig.book.....S, 1965RvPP....1..205B},
\be
\chi \approx 6 \times 10^{-7} T^{5/2} \, \textrm{erg cm$^{-1}$ s$^{-1}$ K$^{-1}$} \,.
\en

(\emph{iii}) The composition of fluid elements can change due to
particle fluxes.  Considering the flux of particles
\be
\bb{Q}_{\rm  c}\equiv-D\b(\b\bcdot\bb{\nabla})c \,,
\en
on the right-hand side of Equation~(\ref{eq:c}) ensures that the diffusion 
of ions is mainly
along magnetic field lines.  This is a good approximation when the
plasma is dilute enough for the ion mean free path to be large
compared to the ion Larmor radius. The concentration $c$ is related to
the mean molecular weight $\mu$ via
\be \frac{1}{\mu} \equiv (1-c)\frac{(1+Z_1)}{\mu_1} + c
\frac{(1+Z_2)}{\mu_2} \,, \en where $\mu_i$ and $Z_i$, with $i=1,2$,
are the molecular weights and the atomic numbers for the two ion
species. The isotropic part of the pressure tensor is thus 
\be 
P= \frac{\rho k_{\rm B} T}{\mu m_{\rm H}} \,,
\en
where $k_{\rm B}$ is the Boltzmann constant and $m_{\rm H}$ is the atomic mass unit.

\subsection{Initial Background State}

We consider a weakly magnetized, plane-parallel atmosphere in a
constant gravitational field $\bb{g}\equiv-g\hat{\bb{z}}$.  The
background magnetic field is weak enough that the mechanical
equilibrium of the atmosphere, with scaleheight $H$, is maintained via
$dP/dz=-g\rho$.  We assume that the medium is stratified in density,
temperature, and composition along the vertical $z$-direction.  In the
equilibrium state, all the particles in the plasma are assumed
to be described by a Maxwellian distribution with the same
temperature, so that $p_\parallel\equiv p_\bot$ initially.

In general, the background heat and particle fluxes do not vanish,
i.e., $\hat{\bb{b}}\bcdot \bb{\nabla} T \ne 0$ and $\hat{\bb{b}}\bcdot
\bb{\nabla} c \ne 0$, unless the magnetic field and the background
gradients are orthogonal. The existence of a well-defined steady
state, i.e., $\bb{\nabla}\bcdot\bb{Q}_{\rm
  s}=\bb{\nabla}\bcdot\bb{Q}_{\rm c}=0$, demands that the background
fluxes should be linear functions of the distance along the
direction of the magnetic field. However, even if this condition is
not strictly satisfied, the dynamics of the modes that we consider is
unlikely to be significantly affected if the local dynamical timescale
is short compared to the timescale in which the entire system evolves
(see also \citealt{2008ApJ...673..758Q}).

\subsection{Validity of the Braginskii-MHD Approximation}

If the pressure anisotropy grows beyond $|p_\parallel-p_\bot|/P \simeq 
\beta^{-1}$, where the plasma $\beta \equiv v_{\rm th}^2/v_{\rm A}^2$,  $v_{\rm
  th}\equiv (2P/\rho)^{1/2}$ is the thermal speed, and $v_{\rm A}\equiv
B/(4\pi\rho)^{1/2}$ is the Alfv{\'e}n speed, the Braginskii-MHD
approximation embodied in Equations~(\ref{eq:rho})--(\ref{eq:S}) becomes 
ill-posed. This is 
because the viscous term introduced in Equation~($\ref{eq:nu_braginskii}$) not
only fails to damp all the kinetic energy available at the viscous
scale but also triggers various fast-growing, micro-scale plasma
instabilities, such as mirror and firehose (see \citealt{2005ApJ...629..139S, 2008PhRvL.100h1301S} 
and references therein). The growth rates of these instabilities are of 
the order of $\gamma \simeq  k_\parallel v_{\rm th}
|p_\parallel-p_\bot|/P$ and thus they can dominate the plasma dynamics
at very small scales if $|p_\parallel-p_\bot|/P \gtrsim
\beta^{-1}$. This poses a challenge in numerical simulations
addressing the non-linear dynamics of the Braginskii-MHD equations
since these instabilities grow formally at the grid scale and some
procedure must be devised in order to capture their effects
(see \citealt{2012ApJ...754..122K} for a detailed discussion).

One possibility, which would prevent the micro-instabilities from
operating at once, is to ignore the effects of pressure anisotropies
and the associated Braginskii viscosity. This was the approach
followed in the seminal papers on the MTI \citep{2001ApJ...562..909B}
and the HBI \citep{2008ApJ...673..758Q}, which showed that both
instabilities grow on the dynamical timescale set by $\od^{-1}
\simeq  (H/g)^{1/2}$.  However, because the timescales involved in
viscous processes are only a factor of a few longer than the dynamical
timescales on which both the HBI and the MTI operate, accounting for small
pressure anisotropies can affect the range of wavenumber over which
these instabilities operate, as well as their growth rates
\citep{2011MNRAS.417..602K}.  Furthermore, the timescales involved in
processes related to ion-diffusion are only a factor of a few larger than
those involved in viscous processes (see below).  Since our aim is to understand the
interplay of the various processes involved in determining the
stability of a medium stratified in both temperature and composition,
we retain the term accounting for Braginskii viscosity,
Equation~($\ref{eq:nu_braginskii}$), in the momentum
Equation~($\ref{eq:v}$).  We argue next that
Equations~(\ref{eq:rho})--(\ref{eq:S}) provide an adequate framework
to analyze the dynamics of small amplitude perturbations of the
stratified atmosphere described.
  
The equilibrium background state over which we perform the stability
analysis is such that $p_\parallel\equiv p_\bot$, and thus there
is an initial period of time for which the pressure anisotropy will
remain small enough that these plasma-micro instabilities can be
ignored. \citet{2012ApJ...754..122K} estimate that the amplitude to which the
fluctuations in the magnetic field can grow before these instabilities
set in, and thus Equations~(\ref{eq:rho})--(\ref{eq:S}) remain
self-consistent, is roughly given by $\delta B_\parallel/B \simeq 
H/(\beta\lambda_{\rm mfp})$, where $\lambda_{\rm mfp}$ stands for the
mean free path between particle collisions.  We can estimate this
value for the ICM as follows.  The plasma $\beta$ increases from $\simeq 
10^2$ in the inner cluster regions to $\simeq  10^4$ in the outer parts,
while the ratio $H/\lambda_{\rm mfp}$ decreases from
$10^3$--$10^2$ in the cluster core to $10^2$--$10$ in the outer region. 
Therefore, the ratio $H/(\beta\lambda_{\rm mfp})$ is larger than unity in 
the central regions of a typical galaxy cluster and decreases outward 
to roughly $10^{-2}$.  We thus conclude that, for the sake of performing 
a linear mode analysis, which is only formally valid when the fluctuations of
all the physical variables are small, e.g., $\delta B/B \ll 1$, the
Braginskii-MHD Equations~(\ref{eq:rho})--(\ref{eq:S}) describes the
problem under consideration self-consistently. These
micro-instabilities are likely to play an important role in the
subsequent non-linear dynamics, but addressing this regime is beyond
the scope of this study.

\section{Stability Analysis}
\label{sec:stability}

\subsection{Linearized Equations}

The modes of interest have associated timescales that are long
compared to the sound crossing time and it thus suffices to work in
the Boussinesq approximation \citep{2001ApJ...562..909B,
  2008ApJ...673..758Q}.  In this limit, the equations for the linear
perturbations $\delta \simeq  e^{\sigma t + i \bb{k}\bcdot\bb{x}}$ become
\footnote{The effects of Braginskii viscosity in the thermal evolution
  of the plasma, which appear in Equation~(\ref{eq:S}) as proportional
  to $(p_\parallel - p_\bot)^2$, are of higher order and thus they do
  not contribute to the linear analysis.}  
\be 
\label{eq:deltav}
\sigma\delta \bb{v}&=&-g\f{\delta\rho}{\rho}\hat{\bb{z}}-i\bb{k}v_{\rm th}^2\left(\f{\delta p_\bot}{P} 
+  \f{1}{\beta}\f{\delta B_\parallel}{B}\right) +ik_\parallel v_{\rm A}^2\f{\delta\bb{B}}{B}\nonumber\\
&&-\b 3k_\parallel^2 \nu_\parallel \delta v_\parallel \,, \\
\label{eq:deltaB}
\sigma\delta \bb{B}&=&ik_\parallel B\delta \bb{v} \,, \\
\label{eq:deltaT}
\sigma\f{\delta\rho}{\rho}&=&\f{N^2}{g}\delta v_z +
\frac{\gamma-1}{\gamma} \kappa k_\parallel^2\f{\delta T}{T}\nonumber\\
&&-i\frac{\gamma-1}{\gamma}\kappa\bb{k}\bcdot\left(\f{d\ln T}{dz}\delta b_z\b
+ b_z\f{d\ln T}{dz}\f{\delta\bb{B}_\bot}{B}\right) \,, \,\,\,\,\,\,\, \\
\label{eq:deltamu}
\sigma\f{\delta\mu}{\mu}&=&-\f{d\ln\mu}{dz}\delta v_z-Dk_\parallel^2\f{\delta \mu}{\mu}\nonumber\\
&&+iD\bb{k}\bcdot\left(\f{d\ln\mu}{dz}\delta b_z\b+b_z\f{d\ln\mu}{dz}\f{\delta\bb{B}_\bot}{B}\right)\,.
\en 
Here, we have defined the anisotropic viscosity coefficient
\be
\nu_\parallel = \frac{1}{2}\frac{v_{\rm th}^2}{\nu_{ii}^{\rm eff}} \,,
\en
where $\nu_{ii}^{\rm eff}$ is an effective collision rate for the binary
mixture (see Appendix~\ref{app:tdi-vs-tvi}), and the thermal diffusion coefficient,
\be
\kappa\equiv \frac{\chi T}{P} \,. 
\en
We have also introduced the Brunt$-$V{\"a}is{\"a}l{\"a} 
frequency, $N^2$, which, in a medium stratified in density, temperature, and 
composition,  is given by
\be
N^2 \equiv \frac{g}{\gamma} \f{d}{dz}\ln P\rho^{-\gamma} = 
g \f{d}{dz}\ln \left(\f{P^{\f{1-\gamma}{\gamma}}T}{\mu} \right)\,.
\en

Note that, in agreement with the Boussinesq approximation, the velocity
perturbations satisfy $\bb{k}\bcdot \delta \bb{v} = 0$ and the
fluctuations in density, temperature, and mean molecular weight are
related via \be \f{\delta \rho}{\rho}+\f{\delta
  T}{T}-\f{\delta\mu}{\mu}=0 \,.
\label{eq:d_rho_T_mu_eq0}
\en

\subsection{Relevant Timescales Across a Mode}

Because of the several physical processes that play a role in the
stability of the dilute atmosphere, the dispersion relation
corresponding to Equations~(\ref{eq:deltav})--(\ref{eq:deltamu}) is
rather involved.  It is thus useful to understand the hierarchy of the
timescales involved in the dynamics of a single Fourier mode in order
to make sensible approximations.  The analysis below applies to the
range of local modes with wavevectors parallel to the magnetic field 
for which the fluid approach is valid, i.e., $H^{-1}<k_\parallel<\lambda_{\rm mfp}^{-1}$,
or
\be
\sqrt{K_{\rm n}} < k_\parallel \sqrt{\lambda_{\rm mfp} H} < \sqrt{K_{\rm n}^{-1}} \,, 
\en
where we have defined the Knudsen number
\be
K_{\rm n} \equiv \frac{\lambda_{\rm mfp}}{H} \,.  
\en

The inverse timescales characterizing the diffusion of heat, momentum, and
particles along magnetic field lines are
\begin{eqnarray}
\tci &\equiv& (\bb{k} \bcdot \hat{\bb{b}})^2 \, \kappa  \, \frac{(\gamma-1)}{\gamma} \,,
\label{eq:tau_conduction} \\
\tvi &\equiv& (\bb{k} \bcdot \hat{\bb{b}})^2 \, 3 \, \nu_\parallel\,, \\
\tdi &\equiv& (\bb{k} \bcdot \hat{\bb{b}})^2 \, D\,.
\end{eqnarray}
For a given mode, the ratio between these timescales is independent of
the direction of the wavevector characterizing the perturbation and
the background magnetic field and is set by plasma processes.  Because
heat conduction is mostly due to electrons, while viscous processes
are dominated by the dynamics of ions, it could be expected that the
associated timescales would satisfy $\tci \gg \tdi$.  However, this is
not the case and a simple estimate leads to $\tci \simeq  6 \tvi$
\citep{2011MNRAS.417..602K}.  
It could be argued that the timescales involved in viscous and diffusion
processes should be of the same order because it is mainly the ion 
dynamics what determines both of them. 
A detailed analysis of the diffusion coefficient for a binary mixture of
ions (see Appendix~\ref{app:tdi-vs-tvi}) shows that  $\tdi \simeq  9 \tvi$ for primordial
composition ($c\simeq 0.25$ or $\mu\simeq  0.6$) and decreases toward 
$\tdi \simeq  3 \tvi$ for the compositions expected at the inner core of 
galaxy clusters according to recent models for helium sedimentation
\citep{2011A&A...533A...6B}.
Since we will be mostly concerned with the two regimes $\tci \gg \od$
or $\od \gg \tci$, as long as the ratio $\tdi/\tvi$ is not too small, 
its particular value will not affect our main conclusions, and we will 
thus consider that $\tdi \simeq  \tvi$.

On the other hand, whether the timescales set by plasma processes are
fast or slow compared to the dynamical timescale $\od^{-1}
\equiv (H/g)^{1/2}$ depends not only on the wavelength of the mode but
also on the direction of the wavevector characterizing the
perturbation with respect to the background magnetic field.
In particular, as shown in \citep{2011MNRAS.417..602K},  
the timescale characterizing conduction across a mode
with parallel wavenumber $k_\parallel$ is related to the dynamical 
timescale via
\begin{eqnarray}
\tci \simeq  10 k_\parallel^2 \lambda_{\rm mfp} H \, \od \,,
\end{eqnarray}
where we have assumed $\gamma = 5/3$ in Equation (\ref{eq:tau_conduction}).
Thus conduction is faster than the dynamical time, i.e.,
$\tci/\od > 1$, if $k_\parallel(\lambda_{\rm mfp}
H)^{1/2} > 1/3$.  If the wavelength of the mode is shorter than this
by a factor of $\tci/\tvi \simeq  6$, e.g., $k_\parallel(\lambda_{\rm
  mfp} H)^{1/2} \gg 1$, then viscous and diffusive processes are also
faster than the dynamical timescale.  Therefore, as a useful
approximate criterion, whether $k_\parallel(\lambda_{\rm mfp}
H)^{1/2}$ is much larger or smaller than unity defines whether the
timescales associated with plasma processes, for that given mode, are
shorter or longer than the dynamical time.  We will thus consider two
different regimes which we refer to as the ``fast" and ``slow"
conduction limit, where the timescales associated with the modes
considered satisfy, respectively,
\begin{eqnarray}
\tci > \tvi \simeq  \tdi \gg \od \quad \textrm{if} \quad k_\parallel \gg (\lambda_{\rm mfp} H)^{-1/2} \,, \quad \\
\od  \gg \tci > \tvi \simeq  \tdi \quad \textrm{if} \quad k_\parallel \ll (\lambda_{\rm mfp} H)^{-1/2} \,. \quad 
\end{eqnarray}

\subsection{The Weak Magnetic Field Limit}

All the timescales related to plasma processes discussed above depend only on the direction
of a given wavevector with respect to the magnetic field. The only time scale that 
depends explicitly on the strength of the field is the one associated with the
Alfv{\'e}n  frequency $\omega_{\rm A} \equiv \bb{k} \bcdot \bb{v}_{\rm A}$.
In order to keep the problem tractable, and given that we are already dealing 
with four different timescales, we will focus on the case where the magnetic field
is so weak that its only physical role is to channel the flux of heat and ions.
The advantage of this limit is that it allows us to address the anisotropic dynamics 
of the weakly collisional magnetized medium without introducing explicitly 
the timescale associated with $\omega_{\rm A}$.

In what follows we focus our attention on modes for which magnetic tension is 
unimportant and thus $\omega_{\rm A} \ll$  min$\{\tci, \od \}$. 
This approximation will be valid for two different ranges of parallel wavenumbers
depending on whether $k_\parallel(\lambda_{\rm mfp} H)^{1/2}$ is much larger 
or smaller than unity.  
For the modes for which conduction is faster than the dynamical timescale, 
i.e., $k_\parallel (\lambda_{\rm mfp} H)^{1/2}\gg 1$, 
we must require $\omega_{\rm A} \ll \od \ll \tci$. 
Using the definitions $v_{\rm th} = (gH)^{1/2}$,
$\beta=v_{\rm th}^2/v_{\rm A}^2$, and $K_n = \lambda_{\rm mfp}/H$, we obtain that 
magnetic tension is negligible provided that
\begin{eqnarray}
1\ll k_\parallel \sqrt{\lambda_{\rm mfp} H} \ll \sqrt{\beta K_n} \,,
\end{eqnarray}
and thus 
\begin{eqnarray}
\omega_{\rm A} \simeq  0 \quad \textrm{for} \quad  \tci \gg \od \quad \textrm{if} \quad \beta K_{\rm n}\gg 1 \,.
\end{eqnarray}
For the modes for which conduction is slow compared to the dynamical timescale, 
i.e., $k_\parallel (\lambda_{\rm mfp} H)^{1/2}\ll1$, 
we must require $\omega_{\rm A} \ll \tci \ll \od$. This is satisfied if
\begin{eqnarray}
\frac{1}{10}\frac{1}{\sqrt{\beta K_n}} \ll k_\parallel \sqrt{\lambda_{\rm mfp} H} \ll 1 \,,
\end{eqnarray}
and therefore
\begin{eqnarray}
\omega_{\rm A} \simeq  0 \quad \textrm{for} \quad  \od \gg \tci \quad \textrm{if} \quad 
\beta K_{\rm n} \gg 10^{-2} \,.
\end{eqnarray}
The plasma $\beta$ ranges from $10^4$ in the outskirts of the ICM
down to $10^2$ in the centers of cool core clusters \citep{2002ARA&A..40..319C}, and
the product $\beta  K_{\rm n}$ ranges from $10^{3}$ in the outskirts of the ICM
decreasing to $10^{-1}$ in the inner regions. Thus the effects of 
magnetic tension can be important in the inner cluster regions.
We address the implications of neglecting magnetic tension when analyzing the
stability of the ICM in further detail in Section \ref{sec:discussion}.

\subsection{General Dispersion Relation for the Dilute, Weakly Magnetized Medium}
\label{sec:modes}

The dispersion relation corresponding to the set of equations for the 
linear perturbations (\ref{eq:deltav})--(\ref{eq:deltamu}) is
\be
\sum_{i=0}^{4} A_i\sigma^{5-i}+\tvi\sum_{i=1}^{5} B_i\sigma^{5-i}=0
\label{eq:fulldisp} \,,
\en
where the coefficients $A_i$ are given by
\begin{eqnarray}
A_0 &\equiv& 1 \,, \\
A_1 &\equiv& \tci  \,, \\ 
A_2 &\equiv& \f{(k_x^2+k_y^2)}{k^2}N^2+\tdi\tci \,, \\
A_3 &\equiv& \tci g\left\{\f{d\ln T}{dz}\f{\K}{k^2}-\f{d\ln\mu}{dz}\f{(k_x^2+k_y^2)}{k^2}\right\} \nonumber \\    &+& \tdi\f{(k_x^2+k_y^2)}{k^2}N^2 \,, \\
A_4 &\equiv& \tdi\tci N_{T/\mu}^2 \f{\mathcal{K}}{k^2} \,,
\end{eqnarray}
while the $B_i$ read
\begin{eqnarray}
B_1 &\equiv& \f{k_\bot^2}{k^2} \,, \\
B_2 &\equiv& \tci \f{k_\bot^2}{k^2} \,, \\
B_3 &\equiv& \tdi\tci\f{k_\bot^2}{k^2}+N^2\f{b_x^2k_y^2}{k^2} \,, \\
B_4 &\equiv& (\tci N_{T/\mu}^2+\tdi N^2) \f{b_x^2k_y^2}{k^2} \,, \\
B_5 &\equiv& \tdi\tci N_{T/\mu}^2 \f{b_x^2k_y^2}{k^2} \,.
\end{eqnarray}
Here we have defined
\begin{eqnarray}
\mathcal{K}&\equiv&(1-2b_z^2)(k_x^2+k_y^2)+2b_xb_zk_xk_z \,, \\
            &=&b_x^2k^2-k_\bot^2+b_x^2k_y^2 \,, \\
            &=&-b_z^2k^2+k_{\parallel}^2+b_x^2k^2_y \,,
\end{eqnarray}
and the two quantities 
\begin{eqnarray}
\label{eq:NTmu}
N_{T\mu}^2 &\equiv& g\f{d}{dz}\ln (T\mu)\,, \\
\label{eq:NT_mu}
N_{T/\mu}^2 &\equiv& g\f{d}{dz}\ln \left(\f{T}{\mu}\right) \,,
\end{eqnarray}
which appear naturally when 
thermal and composition gradients are considered.
The dispersion relation (\ref{eq:fulldisp}) is identical to 
the one derived in \citep{2011MNRAS.417..602K} in the limit 
in which $\omega_{\rm A}$, $d\mu/dz$, and $D$ vanish.
Note that in the limit of a constant composition gradient, i.e., 
$d\mu/dz \rightarrow 0$, both $N_{T\mu}^2$ and $N_{T/\mu}^2$ $\rightarrow g d\ln T/dz$,
which is the logarithmic gradient that plays an important role in the stability
of a homogeneous, dilute, weakly magnetized medium.


\begin{figure*}[t]
\begin{center}
  \includegraphics[width=\textwidth,trim=0 0 0 0]{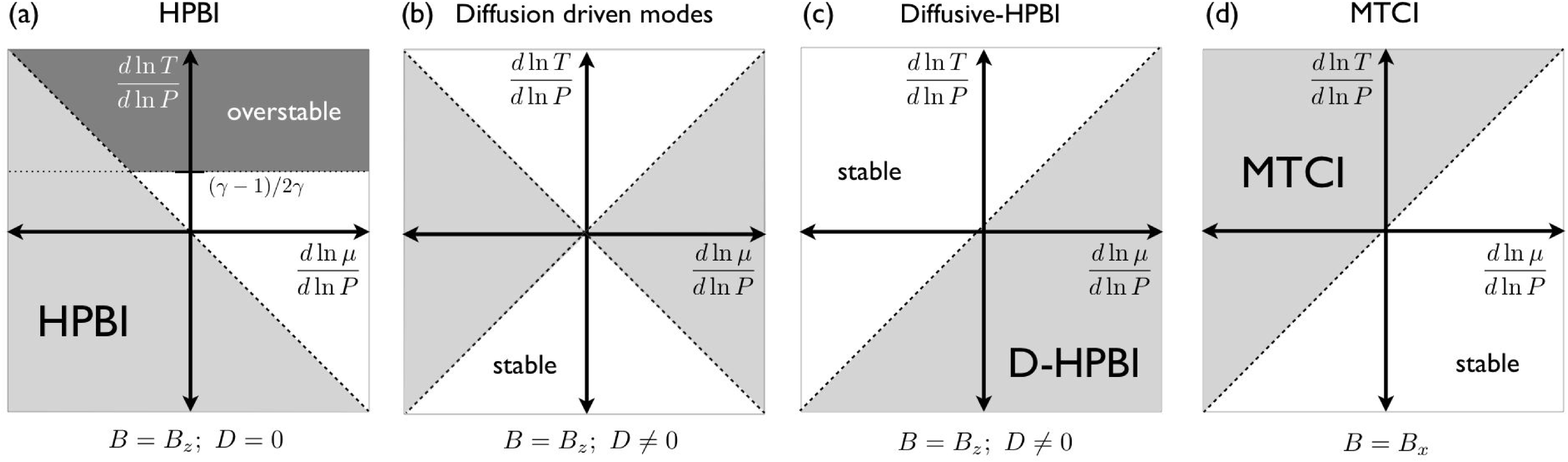}
  \caption{Graphic representation of the stability of modes for which 
    conduction is fast compared to the dynamical timescale, i.e., $\tci \gg \omega_{\rm dyn}$.
    The various panels show the unstable regions (gray) for each of the modes that can be 
    excited when the background magnetic field is parallel ($a$, $b$, and $c$) or
    perpendicular ($d$) to the background thermal and
    composition gradients. The horizontal dotted line in panel (a) represents 
    $d\ln T/d\ln P = (\gamma-1)/2\gamma$; and the dashed lines correspond  
    $d\ln T/d\ln P = \pm d\ln \mu/d\ln P$. 
    Panel (a) shows the region of parameter space which is unstable to the heat- and particle-flux-driven 
    buoyancy instability (HPBI), together with the regions that are overstable to gravity modes.
    If ions can diffuse efficiently along magnetic field lines, i.e., $D\ne0$, there are unstable modes
    that can be driven by diffusion, whether $\od > \tdi \simeq  \tvi$ (b) or $\tdi \simeq  \tvi > \od$ 
    (c). Panel (d) shows the region that is unstable to the magneto-thermo-compositional instability
    (MTCI); in this case, the criterion for instability is insensitive to the value of the diffusion coefficient $D$.  
    }
  \label{fig:fast-conduction}
\end{center}
\end{figure*}

\section{The Fast Conduction Limit}
\label{sec:fast-conduction}

We first consider the stability of the modes for which conduction is faster than the dynamical 
time, i.e., $\tci \gg \od$. This is the regime that corresponds to the well-studied HBI and MTI.

\subsection{Limit of No Ion-Diffusion}

As a first step toward understanding the effects of composition gradients in
the behavior of the HBI and the MTI, we neglect the diffusion of ions along 
magnetic field lines by setting $D=0$. Because we are considering the timescales for ion-diffusion
and viscous processes to be of the same order, i.e., $\tvi \simeq  \tdi$, we also ignore
here the effects of viscosity and set $\nu_\parallel = 0$ for consistency. 
For the modes for which $\tci\gg\od$, the dispersion relation (\ref{eq:fulldisp}) 
yields a (fast) decaying solution, $\sigma\approx-\tci$, together with the two slow modes
\be
\sigma^2\approx -g\left\{\f{d\ln T}{dz}\f{\K}{k^2}-\f{d\ln\mu}{dz}\f{(k_x^2+k_y^2)}{k^2}\right\} \,.
\label{eq:MTIHBI0}
\en
In a homogeneous plasma, these slow modes contain the well-known 
HBI and MTI, depending on the direction of the background magnetic field, i.e., 
\be
\sigma_{\rm HBI}^2&\approx& g\f{d\ln T}{dz}\f{k_\bot^2}{k^2} \quad\quad\quad\quad \textrm{for} \,\,\, b_z=  1 \,,
\label{eq:HBI0} \\
\sigma_{\rm MTI}^2&\approx&-g\f{d\ln T}{dz}\f{k_x^2+k_y^2}{k^2} \quad \textrm{for} \,\,\, b_x=  1 \,.
\label{eq:MTI0}
\en
The conditions for the excitation of the HBI and the MTI are thus
\begin{eqnarray}
\frac{d\ln T}{dz} > 0 \quad \textrm{HBI-unstable} \,,
\label{eq:HBI0-unstable} \\
\frac{d\ln T}{dz} < 0 \quad \textrm{MTI-unstable} \,.
\label{eq:MTI0-unstable}      
\end{eqnarray}
Note that in both cases, the fastest growing modes are those with wavevectors perpendicular to the 
gravitational field.

\subsubsection{Heat- and Particle-Flux Driven Buoyant Instability ($D = 0$)}

Consider a magnetic field parallel to the gravitational field, i.e., $b_z=1$, and 
thus $k_x^2+k_y^2=k_\bot^2$. 
The modified version of the modes that become HBI-unstable in a homogeneous 
medium is given by Equation (\ref{eq:MTIHBI0}). In the medium 
stratified in composition, these modes become
\be
\sigma^2 &\approx& g\frac{d\ln (T\mu)}{dz}\frac{k_\bot^2}{k^2}\,.
\label{eq:sigma_sq-HPBI}
\en
Therefore, neglecting viscous and diffusion processes in a medium
which is stratified in the mean molecular weight leads to modes that
are unstable to a Heat- and Particle-flux-driven Buoyant Instability
(HPBI) if 
\be
\label{eq:HPBI-unstable}
\frac{d\ln T}{dz} > - \frac{d\ln \mu}{dz} \quad \textrm{HPBI-unstable} \,.
\en
The threshold temperature gradient for instability can be negative
if the mean molecular weight increases with height. 

The combination of temperature and composition gradients that is HPBI-unstable is shown 
in panel (a) of Figure~\ref{fig:fast-conduction} where, for the sake 
of convenience, we have defined dimensionless variables $(d\ln \mu/d\ln P, d\ln T/d\ln P)$
in terms of the logarithmic pressure gradient $d\ln P/dz \equiv -1/H$.
 
\subsubsection{Overstable Modes ($D = 0$)}

\citet{2010ApJ...720L..97B} showed that temperature gradients that are stable 
to the HBI can nevertheless be subject to overstable gravity modes. This result 
can be extended to include non-vanishing composition gradients, i.e., 
there is a range of modes with $\tci>\od$ and $\sigma\simeq \od$ that 
can become overstable when the heat- and particle-flux-driven buoyancy 
instability (HPBI) does not operate.  Calculating these modes requires 
retaining higher order terms in the dispersion relation, which becomes
\be
\sigma^3+\tci\sigma^2+\f{k_\bot^2}{k^2}N^2\sigma-\tci N^2_{T\mu}\f{k_\bot^2}{k^2}=0 \,.
\en
In this regime, we can treat the first and third terms on the left-hand 
side as perturbations and extend the solutions (\ref{eq:sigma_sq-HPBI}) to contain 
corrections of order $\od^2/\tci$:
\be
\sigma\approx\pm i\sqrt{-N^2_{T\mu}\f{k_\bot^2}{k^2}}-\f{N^2+N^2_{T\mu}}{2\tci} \,.
\label{eq:OSHPBI}
\en
Therefore, modes that are stable according to the HPBI-stability criterion, i.e.,
$N^2_{T\mu} < 0$ (Equation~[\ref{eq:HPBI-unstable}]), can become overstable if 
$N^2_{T\mu}< - N^2$. In terms of the dimensionless
variables introduced earlier, these requirements become
\be
\label{eq:OHBI-overstable}
\f{d\ln T}{d\ln P} > \textrm{max} \left\{- \f{d\ln \mu}{d\ln P}, \f{\gamma -1}{2\gamma} \right\} \quad \textrm{HPBI-overstable} \,. \,\,\,
\en
The combination of temperature and composition gradients that is subject to overstability
is shown in dark gray in panel (a) of Figure~\ref{fig:fast-conduction}.

\subsubsection{Magneto-Thermo-Compositional Instability ($D = 0$)}

In order to understand how the MTI is modified in the presence of composition gradients
we consider a horizontal magnetic field along the $x$-axis, i.e., $b_x=  1$ and 
focus on the modes for which $\tci \gg \omega_{\rm dyn}$. 
In the presence of a gradient in the mean molecular weight, Equation (\ref{eq:MTIHBI0}) 
gives the generalization of the modes that become MTI-unstable 
\be
\sigma^2& \approx& -g\f{d\ln (T/\mu)}{dz}\f{k_x^2+k_y^2}{k^2}\,.
\label{eq:sigma_sq-MTCI}
\en
Thus a non-vanishing gradient in the mean molecular weight sets an
upper bound for the temperature gradients that are 
magneto-thermo-compositional instability (MTCI)-unstable
\be
\label{eq:MTCI-unstable}
\f{d\ln T}{dz} < \f{d\ln \mu}{dz} \quad \textrm{MTCI-unstable} \,.
\en
Panel (d) in Figure~\ref{fig:fast-conduction} shows a graphical 
representation of the region of parameter space that is subject to 
unstable MTCI modes in a medium that is stratified in composition and 
temperature.

\subsection{Ion-Diffusion Along Magnetic Field Lines}

We now analyze the effects of including ion-diffusion induced by 
the background composition gradients. Since, for a given mode, 
viscous and diffusion timescales are of the same order, i.e., 
$\tvi\simeq \tdi$, we also consider the effects of anisotropic viscosity for consistency.

\subsubsection{Heat-and Particle-Flux-Driven Buoyant Instability ($D\ne0$)}

Starting from the general dispersion relation in the case where $b_z= 1$,
it can be seen that, if we consider modes for which conduction is faster than any 
other timescale, there is a fast decaying solution $\sigma=-\tci$ which can be used to 
self-consistently obtain three more modes satisfying
\begin{eqnarray}
\sigma^3+\left(\tdi+\tvi\f{k_\bot^2}{k^2}\right)\sigma^2&+&
\f{k_\bot^2}{k^2}\left(\tdi\tvi-N^2_{T\mu}\right)\sigma \nonumber\\
&-&\tdi N^2_{T/\mu}\f{k_\bot^2}{k^2}=0 \,. \quad
\end{eqnarray}
However, the leading order solutions to this dispersion relation depend on 
whether the dynamical timescale is fast or slow with respect to 
the diffusion and viscous timescales across the mode. 
We thus need to consider these two cases separately.

\textit{Slow diffusion}.
The modes for which $\od>\tvi\simeq \tdi$ contain the generalization 
of the HBI for non-vanishing composition gradients\footnote{When a finite diffusion 
timescale is considered, i.e., $D\ne 0$, the term that contributes to the 
generalization of the HBI is the second term inside brackets in the right 
hand side of Equation~(\ref{eq:deltamu}).}
\begin{eqnarray}
\sigma^2\approx N^2_{T\mu}\f{k_\bot^2}{k^2} \,,
\label{eq:sigma_sq-HBI}
\end{eqnarray}
which grows dynamically if $N^2_{T\mu} >0$, or $dT/dz>-d\mu/dz$.  
This is the same condition for the onset of the HPBI in 
Equation~(\ref{eq:HPBI-unstable}) in the absence of diffusion.
It is worth mentioning that even if the combination of temperature and 
composition gradients is such that $N^2_{T\mu}<0$ (and thus the plasma is 
HPBI-stable), overstable modes can be excited just as in the non-diffusive 
case. In fact, the condition for overstability can be shown to be exactly 
similar to (\ref{eq:OHBI-overstable}), provided that $\od> {\rm max}\{\tci\tvi,\tci\tdi\}$.
There also exists a slower mode  driven by ion-diffusion
\begin{eqnarray}
\sigma\approx-\tdi\f{N^2_{T/\mu}}{N^2_{T\mu}} \,,
\end{eqnarray}
which grows in the region of parameter space where
$N^2_{T/\mu}/N^2_{T\mu}<0$ or, in terms of the background gradients,
wherever $d\ln T/d\ln P < \left|d\ln \mu/d\ln P\right|$,  
as it is shown in panel (b) of Figure~\ref{fig:fast-conduction}.  
Note that these modes can grow on a diffusion timescale even if the system is HPBI stable,
i.e., $N^2_{T\mu} <0$, provided that $N^2_{T/\mu} > 0$.  As we show
below, this latter requirement becomes the deciding one for those
modes for which diffusion is not slow compared to the dynamical time.

\textit{Fast diffusion}. For the modes for which $\tvi\simeq \tdi>\od$,
there are two solutions that decay on the diffusion and viscous timescales,
i.e., $\sigma\approx-\tdi$ and $\sigma \approx -\tvi k_\bot^2/k^2$, and
a third one 
\be 
\sigma \approx \f{N^2_{T/\mu}}{\tvi} \,, 
\label{eq:sigma_sq-DHPBI}
\en 
which can become unstable if $N^2_{T/\mu} > 0$. 
This region of parameter space in temperature and composition
gradients is unstable to a diffusive version of the
Heat- and Particle-flux-driven Buoyant Instability (D-HPBI)
\be 
\f{d\ln T}{dz} > \f{d\ln \mu}{dz} \quad \textrm{D-HPBI-unstable} \,, 
\en 
and it is shown in panel (c) of Figure~\ref{fig:fast-conduction}.
 
In summary, we conclude that when finite viscous and diffusion timescales 
are considered, there can be unstable modes driven by diffusion (with $\od>\tdi$) 
even if the temperature and the composition gradients do not satisfy inequality 
(\ref{eq:HPBI-unstable}), i.e., they are stable to the HPBI
in the absence of diffusion, provided that $d\ln T > d\ln \mu$.
Note that this is the very requirement for the existence of unstable modified HBI modes
when finite diffusion timescales are relevant (with $\tdi>\od$) and differs
from the condition  (\ref{eq:HPBI-unstable}). We shall discuss the physical 
reason behind this change in condition of instability in Section~\ref{sec:physics}.

\subsubsection{Magneto-Thermo-Compositional Instability ($D\ne0$)}

In order to understand how ion-diffusion driven by a composition gradient
affects the MTI, we consider $D\ne0$ and modes for which $\tci \gg \omega_{\rm dyn}$
when $b_x=  1$.
In this case, the dispersion relation factorizes and leads to a fast decaying 
solution $\sigma=-\tci$ together with 
\be
\sigma^3+\tvi\f{k_\bot^2}{k^2}\sigma^2+N^2_{T/\mu}\f{k_x^2+k_y^2}{k^2}\sigma+\tvi N^2_{T/\mu}\f{k^2_y}{k^2}=0 \,. \qquad
\label{eq:MTIkyne0}
\en

For the modes satisfying $\od>\tvi$, the three solutions to this cubic equation
correspond to a decaying viscous mode 
\be
\sigma\approx-\tvi\f{k_y^2}{k_x^2+k_y^2} \,,
\en
and a pair of roots that contain the generalization of the MTI in the presence 
of a mean molecular weight gradient and $D\ne0$
\be
\sigma^2\approx - N^2_{T/\mu} \f{k^2_x+k^2_y} {k^2} \,,
\label{eq:sigma_sq-DMTCI}
\en
which grows dynamically if $N^2_{T/\mu} <0$, or $dT/dz>d\mu/dz$.

This is the same condition for the onset of the MTCI in 
Equation~(\ref{eq:MTCI-unstable}) in the absence of diffusion.
As we found before,  a non-vanishing gradient in the mean molecular weight sets 
a upper bound for the temperature gradients that are MTI-unstable.
This temperature gradient \emph{can} be positive when the
composition gradient is negative (i.e., mean molecular weight
increasing with height).  Note that this statement is independent of
the value of the diffusion coefficient $D$, i.e., and thus of whether
or not ions diffuse effectively along magnetic field lines on a
dynamical timescale.  

For the modes such that $\tvi>\od$, the three solutions to Equation~(\ref{eq:MTIkyne0})
correspond to another viscously damped mode
\be
\sigma\approx-\tvi\f{k_\bot^2}{k^2} \,,
\en
together with a generalization of the modes identified as Alfv{\'e}nic-MTI in 
\citet{2011MNRAS.417..602K}, i.e., 
\be
\sigma\approx -\f{N^2_{T/\mu}}{2\tvi}\f{k_x^2}{k_\bot^2}\pm i \sqrt{N^2_{T/\mu}}\f{k_y}{k_\bot} \,.
\en
These modes grow dynamically if $N^2_{T/\mu} < 0$, which corresponds
again to the combination of temperature and composition gradients
satisfying inequality (\ref{eq:MTCI-unstable}).


\begin{figure*}[t]
\begin{center}
  \includegraphics[width=\textwidth,trim=0 0 0 0]{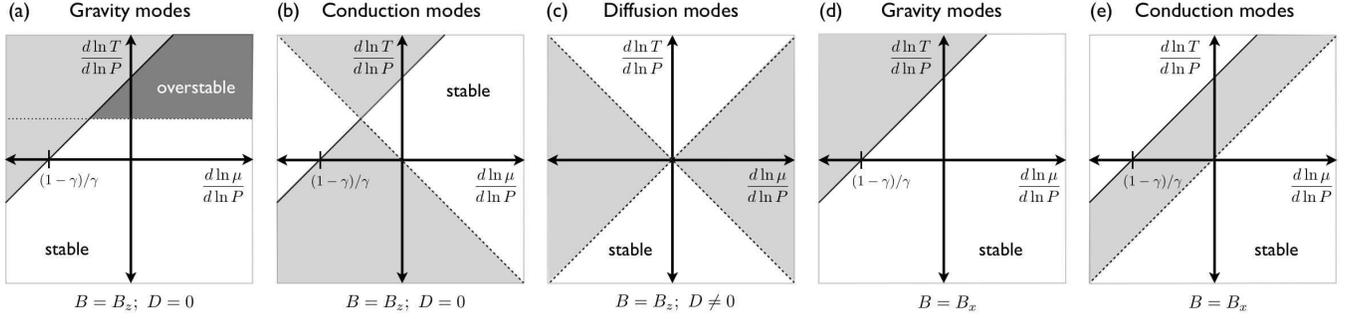}
  \caption{Graphic representation of the stability of modes for which
    conduction is slow compared to the dynamical timescale, i.e.,
    $\omega_{\rm dyn} \gg \tci$.  The various panels show the unstable
    regions (gray) for each of the modes that can be excited when the
    background magnetic field is parallel ($a$, $b$, and $c$) or
    perpendicular ($d$ and $e$) to the background thermal and
    composition gradients.  The solid line corresponds to $N^2=0$; the
    horizontal dotted line represents $d\ln T/d\ln P =
    (\gamma-1)/2\gamma$; and the dashed lines correspond to $d\ln
    T/d\ln P = \pm d\ln \mu/d\ln P$.  For $B= B_z$, gravity modes
    ($a$) can become either stable or overstable, while modes driven
    by conduction can become unstable ($b$). If ions can diffuse
    efficiently along magnetic field lines, a new type of mode can
    become unstable ($c$).  For $B= B_x$, both gravity modes ($d$) and
    conduction modes can become unstable ($e$); while ion-diffusion
    only leads to decaying modes.}
  \label{fig:slow-conduction}
\end{center}
\end{figure*}

\section{The Slow Conduction Limit}
\label{sec:slow-conduction}

We now analyze the stability of the modes whose associated timescales 
satisfy $\omega_{\rm dyn} \gg \tci > \tvi \simeq  \tdi$.  For a 
homogeneous plasma, these modes encompass the overstable $g$-modes
studied in \citet{2010ApJ...720L..97B}.  The set of
Equations~(\ref{eq:deltav})--(\ref{eq:d_rho_T_mu_eq0}) allows us to
address the behavior of these, as well as other new modes, in the
presence of a non-vanishing gradient in the mean molecular weight and
account self-consistently for the diffusion of ions along magnetic
field lines.

\subsection{Heat and Ion Diffusion Along Vertical Magnetic Fields}
\label{sec:Bz_Dneq0}

In the case where the background magnetic field is parallel to the gravitational
field, $b_z=  1$, the dispersion relation (\ref{eq:fulldisp}) reduces to 
\be
\sigma^4+a_1\sigma^3+a_2\sigma^2+a_3\sigma+a_4=0 \,,
\label{eq:pol_a} 
\en
with coefficients 
\begin{eqnarray}
a_1 &=& \tci + \tdi + \f{k_\bot^2}{k^2}\tvi \approx \tci \,, \nonumber \\
\label{eq:a1}
a_2 &=&\f{k_\bot^2}{k^2}N^2+ \tci\left(\tdi + \f{k_\bot^2}{k^2}\tvi\right) 
\approx \f{k_\bot^2}{k^2}N^2 \,, \\
a_3 &=&  \f{k_\bot^2}{k^2} \left[ - \tci N_{T\mu}^2  + {\tdi} (N^2 + \tci\tvi) \right] \nonumber  \\
&\approx& - \tci \f{k_\bot^2}{k^2} N_{T\mu} ^2 \,,\,\,\,\,\,\,\,\,\,\,\\
a_4 &=& - {\tdi} \tci \f{k_\bot^2}{k^2}N_{T/\mu}^2  \,. 
\label{eq:a4}
\end{eqnarray}

The Routh--Hurwitz stability criteria that predict exclusively
negative real parts for the roots of the quartic polynomial with
real coefficients require $a_1>0, a_1a_2-a_3>0$, 
$a_1a_2a_3-a_1^2a_4-a_3^2>0$, and $a_4>0$.  
The first of these conditions is trivially satisfied,
while the other three imply, respectively,
\begin{eqnarray}
\label{eq:RH-3.2}
N_{T\mu}^2 &<& 0 \,,\\
\label{eq:RH-3.3}
N^2 +N_{T\mu}^2 &>& 0 \,, \\
\label{eq:RH-4.2}
N^2_{T/\mu} &<& 0 \,.
\end{eqnarray}
In the absence of diffusion, only the conditions (\ref{eq:RH-3.2}) and (\ref{eq:RH-3.3})
need to be met in order to ensure stability, while the condition (\ref{eq:RH-4.2}) should also 
be required for finite diffusion timescales.

To leading order, two of the solutions of Equation~(\ref{eq:pol_a})
are given by
$\sigma\approx\pm i a_2^{1/2}+(a_3-a_1a_2)/2a_2$, i.e.,
\be
\sigma&\approx&\pm i\f{k_\bot}{k}\sqrt{N^2} - \f{\tci}{2} \left(1 + \f{N_{T\mu}^2}{N^2}\right) \,,
\label{eq:fast_overstable_Bz}
\en
which correspond to gravity modes. In the absence of a gradient in the mean molecular
weight, these reduce to the $g$-modes discussed in \citet{2010ApJ...720L..97B}.
The third root is given by
$\sigma\approx -a_3/a_2$, i.e.,
\be
\sigma&\approx& \tci \frac{N^2_{T\mu}}{N^2} \,,
\label{eq:conduction_Bz}
\en
and corresponds to a mode driven by conduction.
Assuming that $N^2>0$, $g$-modes are overstable if the condition 
(\ref{eq:RH-3.3}) is not satisfied, while conduction modes are unstable if
Equation~(\ref{eq:RH-3.2}) is not fulfilled. 
The fourth solution consists of a mode driven by ion-diffusion
\be
\sigma\approx-\tau^{-1}_{\rm d}\frac{N_{T/\mu}^2}{N_{T\mu}^2} \,,
\label{eq:diffusion}
\en
which is unstable if either criterion 
(\ref{eq:RH-3.2}) or (\ref{eq:RH-4.2}) is unfulfilled.

It is useful to understand what types of modes can be excited in 
the different regions of the parameter space spanned by the gradients in
temperature and composition. The Routh--Hurwitz stability criteria take
simple forms when expressed in terms of the dimensionless gradients defined
in terms of the pressure. The classical requirement for stability against 
buoyancy, i.e., $N^2 > 0$ becomes
\be
\label{eq:TP-cond-1}
\frac{d\ln T}{d\ln P} < \frac{d\ln \mu}{d\ln P} + \frac{\gamma - 1}{\gamma} \,,
\en 
while the conditions (\ref{eq:RH-3.2}) and (\ref{eq:RH-3.3}) become, respectively,
\begin{eqnarray}
\label{eq:TP-cond-2}
\frac{d\ln T}{d\ln P} &>& -\frac{d\ln \mu}{d\ln P} \,,  \\
\label{eq:TP-cond-3}
\frac{d\ln T}{d\ln P} &<& \frac{\gamma - 1}{2\gamma} \,.
\end{eqnarray}
If ions can diffuse along magnetic field lines, in addition to
requiring that the gradients in temperature, pressure, and mean molecular 
weight satisfy the inequalities (\ref{eq:TP-cond-2}) and (\ref{eq:TP-cond-3}), 
the inequality  (\ref{eq:RH-4.2}) must also be satisfied, i.e.,
\begin{eqnarray}
\frac{d\ln T}{d\ln P} &>& \frac{d\ln \mu}{d\ln P} \,.
\label{eq:TP-cond-4}
\end{eqnarray}

Panel (a) in Figure~\ref{fig:slow-conduction} shows the
regions of parameter space where $g$-modes in
Equation~(\ref{eq:fast_overstable_Bz}) are stable, overstable, or
unstable (gray area).  Panel (b) shows that the modes in
Equation~(\ref{eq:conduction_Bz}), which are driven by conduction, can
be either stable or unstable. Note that in the region of parameter
space where both gravity and conduction modes are overstable/unstable
they both grow with comparable rates.  
Panel (c) in Figure~\ref{fig:slow-conduction} shows that the modes in
Equation~(\ref{eq:diffusion}), which are driven by diffusion, can be
either stable or unstable (gray). Their growth rates are estimated to
be an order of magnitude smaller than either $g$-modes or conduction
modes. The importance of these diffusion modes resides in that they
can become unstable in regions of parameter space which are stable
against $g$-modes and conduction modes.

\subsection{Heat and Ion Diffusion Along Horizontal Magnetic Fields}
\label{sec:Bx_all}

If the background magnetic field is perpendicular to the thermal and the composition
gradients, i.e., $b_x=  1$, the dispersion relation (\ref{eq:fulldisp}) becomes
\footnote{In the absence of diffusion, the only result that is modified in this 
section is that the root $\sigma=-\tdi$ for $D\ne 0$ becomes $\sigma=0$ for $D=0$.} 
\be
\sigma^4+b_1\sigma^3+b_2\sigma^2+b_3\sigma+b_4=0 \,,
\label{eq:pol_c}
\en
where the coefficients
\be
b_1 &\approx& \tci+\tvi\f{k^2_\bot}{k^2}\,,\\
b_2 &\approx& N^2 \f{k^2_x+k_y^2}{k^2}+\tci\tvi\f{k^2_\bot}{k^2}\,,\\
b_3 &\approx& \tci N^2_{T/\mu}\f{k^2_x+k_y^2}{k^2}+\tvi N^2\f{k_y^2}{k^2}\,,\\
b_4 &\approx& \tci\tvi N^2_{T/\mu}\f{k_y^2}{k^2}\,,
\en
are subject to the same considerations employed in deriving the approximate 
expressions for the coefficients $a_i$. 

The Routh--Hurwitz stability criteria require
$b_1>0$, $b_1b_2-b_3>0$, $b_1b_2b_3-b_1^2b_4-b_3^2>$, and  $b_4>0$.
The first condition is trivially satisfied, while, 
in the limit under consideration, i.e., $\omega_{\rm dyn}\gg \tci > \tvi$, 
the other three conditions become, respectively,
\be
&& N^2-N^2_{T/\mu}>0\,,
\label{eq:cond_c2}
\\
&& N^2_{T/\mu}(N^2-N^2_{T/\mu})>0\,,\\
&&N^2_{T/\mu}>0 \,.
\label{eq:c5>0}
\en 
The inequality (\ref{eq:cond_c2}) is always satisfied, since it can be written as 
\be
\label{eq:dPdz_sq}
\frac{\gamma-1}{\gamma P\rho} \left(\frac{dP}{dz}\right)^2>0 \,.
\en
Therefore, the only independent condition required for stability is $N^2_{T/\mu}>0$,
or $d\ln T/dz > d \ln \mu/dz$.

Two of the approximate solutions to the dispersion relation (\ref{eq:pol_c}) are given by
$\sigma\approx\pm i b_2^{1/2}+(b_3-b_1b_2)/2b_2$, i.e.,
\be
\sigma\approx\pm i\f{\sqrt{k_x^2+k_y^2}}{k}\sqrt{N^2}-\f{\tci}{2}\left(1-\f{N^2_{T/\mu}}{N^2}\right) \,.
\label{eq:fast_overstable_Bx}
\en
These correspond to gravity modes, which cannot become overstable on account of 
Equation~(\ref{eq:dPdz_sq}). The other two solutions correspond to a conduction
and a viscous (decaying) mode, which are, respectively,
\be
\sigma\approx-\tci \frac{N^2_{T/\mu}}{N^2} \,,  \qquad
\sigma\approx-\tvi \frac{k_y^2}{k_x^2+k_y^2} \,.
\label{eq:conduction_Bx}
\en

We conclude that when the magnetic field is perpendicular to 
the temperature and the composition gradients, the stability of $g$-modes 
requires only that $N^2>0$, whereas the stability of conduction modes requires also
\be
\frac{d\ln T}{d\ln P} < \frac{d\ln \mu}{d\ln P}  \,.
\label{eq:TP-cond-5}
\en 
This is illustrated in panel (d) in Figure~\ref{fig:slow-conduction}, 
which shows the regions of parameter space where $g$-modes in
Equation~(\ref{eq:fast_overstable_Bx}) are stable or unstable (gray).
Panel (e) shows that the modes in
Equation~(\ref{eq:conduction_Bx}), which are driven by conduction, can
be either stable or unstable. These regions are significantly
different from the corresponding regions in panel (b), for which the
direction of the background magnetic field is parallel to the
direction of the temperature and the composition gradients.  Note that,
unlike the case where $b_z = 1$, $g$-modes and conduction modes cannot
be simultaneously unstable when $b_x= 1$.  


\begin{figure*}[t]
\begin{center}
  \includegraphics[width=\textwidth,trim=0 0 0 0]{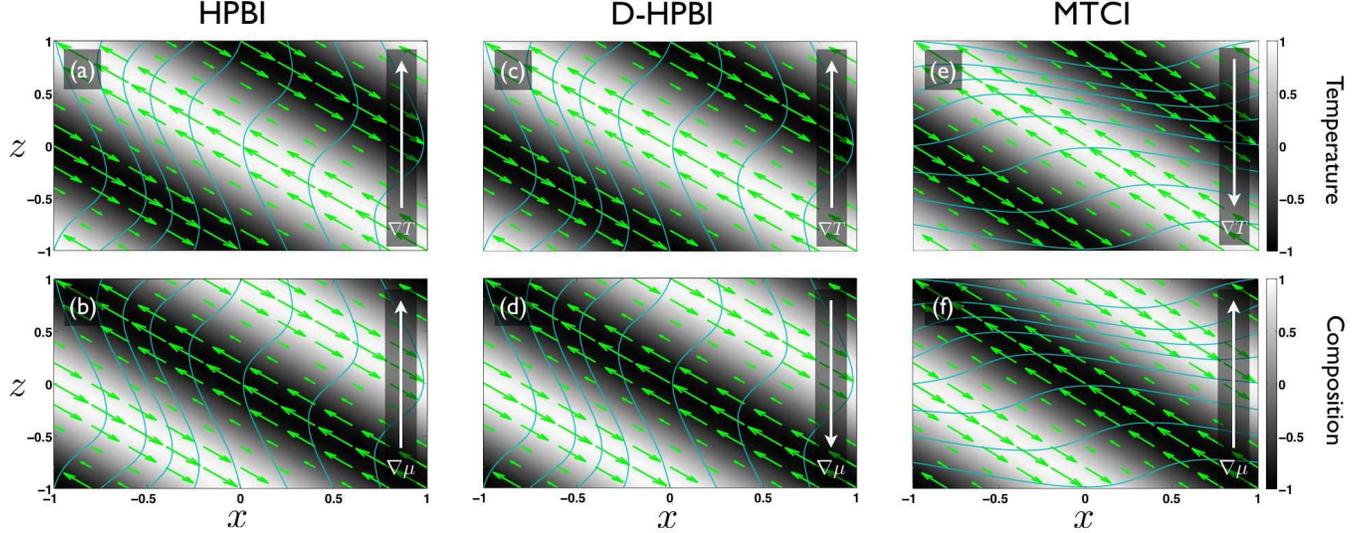}
  \caption{
  Schematic representation of various unstable modes in a weakly magnetized plasma
  with temperature and composition gradients parallel to the gravitational field 
  $\bb{g}=-g \hat{\bb z}$. The three sets of panels (a,b); (c,d); and (e,f) show the 
  modes that are unstable to the heat- and particle-flux-driven buoyancy instability (HPBI), 
  the diffusive-HPBI, and the magneto-thermo-compositional instability (MTCI), respectively.
  The arrows represent the Lagrangian displacements, assumed to be of the form $\bb{\xi} = 
  \bb{\xi}_0 \cos(k_x x+k_z z)$, with $k_x=k_z$. The continuous lines represent the magnetic 
  field lines, which are assumed to be parallel (HPBI, D-HPBI) or perpendicular (MTCI) to the 
  gravitational field in the equilibrium state. The gray-scale contours show the temperature 
  and the mean molecular weight fluctuations relative to the background gradients, which are 
  shown with arrows indicating their directions in each of the cases considered. 
}
  \label{fig:physics}
\end{center}
\end{figure*}

\section{The Physics Driving Unstable Modes}
\label{sec:physics}

In previous sections we have seen that the criteria for instability result 
from an intricate interplay between the thermal and the composition gradients. 
The final expressions for the inequalities that must be satisfied for the onset of 
unstable modes depend implicitly on whether conduction across the associated scale
is faster or slower than the dynamical timescale, the direction of the background field,
and the ability of the ions to diffuse along magnetic field lines.
We now analyze in detail the eigenmodes corresponding to some 
of the most relevant instabilities and shed light on the physical phenomena 
that play a role in determining the stability of the plasma.
This exercise is similar in spirit to the ones presented in 
\citet{2008ApJ...673..758Q} and \citet{2001ApJ...562..909B} to highlight the 
physics of the HBI and the MTI but emphasizes the effects of composition 
gradients and ion-diffusion along magnetic field lines. 
For the sake of simplicity we assume that the perturbations under 
consideration correspond to modes with $k_y=0$.
In this case, the components of the Lagrangian displacement, 
$\partial \bb{\xi}/\partial t = \delta \bb{v}$, are related 
via $\xi_x=-(k_z/k_x)\xi_z$.

\subsection{Heat- and Particle-Flux Driven Buoyancy Instability}

Let us first consider a background  magnetic field with $b_z=1$
and focus on perturbations with wavelengths such that the associated 
timescales satisfy $\tci\gg\omega_{\rm dyn}\gg\tvi\simeq \tdi$. For these modes,
ion-diffusion along the magnetic field lines is inefficient.
In order to understand the effect of a composition gradient, we retain the 
dominant terms in Equations~(\ref{eq:deltaT}) and (\ref{eq:deltamu}), which leads to
\be
\delta T=\frac{d T}{dz}\xi_z \,,  \qquad \delta\mu=-\frac{d\mu}{dz}\xi_z  \,.
\en
Because the fluctuations in density, temperature, and mean molecular weight are related via 
Equation~(\ref{eq:d_rho_T_mu_eq0}), this implies that the relative change in 
density of a fluid element which is vertically displaced by $\xi_z$ is given by 
\be
\label{eq:delta_rho_HPBI}
\f{\delta\rho}{\rho}=  -\left(\frac{d\ln T}{dz} + \frac{d\ln\mu}{dz}\right) \xi_z  \qquad \textrm{HPBI}\,.
\en
It is easier to understand the physics behind this equation by analyzing first 
the effects of each of the two terms on the right-hand side separately. 

In a homogeneous medium, $d\mu/dz = 0$, the term proportional to the temperature 
gradient is responsible for the onset of the HBI when $dT/dz >0$. This is illustrated in 
panel (a) in Figure~\ref{fig:physics}, which shows that a fluid element that is 
displaced with  $\xi_z >0$ ($\xi_z <0$) is effectively heated (cooled) by the increased (decrease) flux 
of heat due to the convergence (divergence) of field lines that results from the displacement 
of the fluid element. This causes the fluid element to expand (contract), and thus
attain a density which is lower (higher) than the surrounding medium. This
leads to the runaway process known as the HBI.

An isothermal environment, $dT/dz = 0$, stratified in composition is unstable if
$d\mu/dz > 0$, as shown in panel (b) in Figure~\ref{fig:physics}.
Because the diffusion of ions along magnetic field lines is inefficient, 
the mean molecular weight of a fluid element that is displaced with $\xi_z >0$ ($\xi_z <0$) 
is lower (higher) than the surrounding medium. This results in a displaced fluid element
with a density which is lower (higher) than the density of the surrounding medium, which will 
thus rise (sink). 

If the temperature and the composition gradients are both 
positive (negative) then these arguments act in consonance and lead to the conclusion 
that the plasma is unstable (stable). On the other hand, if the temperature and the composition 
gradients have different signs it follows that if $dT/dz >0$ ($d\mu/dz > 0$) is steep enough 
then the expansion induced in an upwardly displaced fluid element can offset the stabilizing 
effects of $d\ln \mu/dz < 0$ ($d\ln T/dz < 0$) and the plasma will be unstable, giving rise
to an HPBI.

The arguments outlined here can also be derived from the equation of motions for the 
Lagrangian displacement $\xi_z$.  The buoyancy force per unit volume 
on a vertically displaced fluid element produces an acceleration given by
\be
\frac{d^2\xi_z}{dt^2}=-g\frac{\delta \rho}{\rho} \,,
\en
and thus, according to Equation~(\ref{eq:delta_rho_HPBI}),
\be
\frac{d^2\xi_z}{dt^2} &=&  N_{T\mu}^2  \, \xi_z \,.
\en
This leads to an instability if $N_{T\mu}^2>0$, in agreement with the results of
Section~\ref{sec:fast-conduction}.

\subsection{Diffusive Heat- and Particle-Flux Driven Buoyancy Instability}

The effects of ion-diffusion are not negligible for the modes for which
the associated timescales satisfy $\tci>\tvi\simeq \tdi\gg\omega_{\rm dyn}$.
To leading order, 
the changes in the temperature and the mean molecular weight in a fluid element 
which is vertically displaced by $\xi_z$ in a background magnetic field 
with $b_z=1$ are given by
\be
\delta T=\frac{d T}{dz}\xi_z \,,  \qquad \delta\mu=\frac{d\mu}{dz}\xi_z  \,.
\en
Note that because the dominant terms in Equation~(\ref{eq:deltamu}) 
are both proportional to $D$, the fractional change in the mean molecular weight is 
independent of the value of the diffusion coefficient. The fractional change in density is thus
\be
\label{eq:delta_rho_D-HPBI}
\f{\delta\rho}{\rho}=  -\left(\frac{d\ln T}{dz} - \frac{d\ln\mu}{dz}\right) \xi_z \qquad \textrm{D-HPBI}\,.
\en
The role played by the background temperature gradient is identical to
the one discussed in the absence of ion-diffusion, and this situation is 
shown for the sake of clarity in panel (c) in Figure~\ref{fig:physics}. 
However, the contribution from the mean molecular weight gradient is now 
the opposite. This difference in sign is due to the manifest role played by 
ion-diffusion as an effective agent to tap into the free energy available 
in the background particle flux needed to maintain the composition gradient. 

In order to understand the role played by ion-diffusion let us focus on a background
with constant temperature, $dT/dz = 0$. The term proportional to the composition
gradient is responsible for the onset of the D-HPBI when $d\mu/dz <0$. This is illustrated in 
panel (d) in  Figure~\ref{fig:physics}, which shows that the mean molecular weight 
of a fluid element that is displaced with  $\xi_z >0$ ($\xi_z <0$) will decrease (increase) 
because of  the decrease (increase) in the flux of particles out of (into) it due to the diverge 
(convergence) of field lines that results from the displacement of the fluid parcel. This 
causes the density of the fluid element to decrease (increase) with respect to the surrounding 
medium, leading to a runaway process. A background temperature gradient with $dT/dz > 0$ 
will reinforce this process leading to enhanced buoyancy.
In general, when both the temperature and the composition gradients are non-zero, 
the condition for D-HPBI is $d\ln T/dz<d\ln \mu/dz$.
This is reflected in the equation of motion for the Lagrangian displacement
\be
\frac{d^2\xi_z}{dt^2} &=&  N_{T/\mu}^2  \, \xi_z \,, 
\en
which has exponentially growing solutions when $N_{T/\mu}^2>0$, as we have seen before.

\subsection{Magneto-Thermo-Compositional Instability}

Let us now consider a background horizontal magnetic field with $b_x=1$
and modes for which conduction is faster than any other timescale.
When a fluid element is displaced by $\xi_z$ from its equilibrium position, 
the fluctuations in temperature and composition are given by
\be
\delta T=-\frac{d T}{dz}\xi_z \,,  \qquad \delta\mu=-\frac{d\mu}{dz}\xi_z  \,.
\en
It is important to note that this relationship between the fractional change in
the mean molecular weight and the Lagrangian displacement holds regardless of whether
the dynamical timescale is fast or slow with respect to the viscous and
diffusion timescales. The only difference is that in the former case
the leading order terms in Equation~(\ref{eq:deltamu}) that lead to 
$\delta\mu=- (d\mu/dz)\xi_z$ are independent of the coefficient $D$, 
while in the opposite limit, the dominant terms are those proportional to $D$.
In either case, the relative change in density becomes
\be
\label{eq:delta_rho_MTCI}
\f{\delta\rho}{\rho}=  \left(\frac{d\ln T}{dz} - \frac{d\ln\mu}{dz}\right) 
\xi_z \qquad \textrm{MTCI} \,.
\en

In this case, the term proportional to the temperature 
gradient is responsible for the onset of the MTI when $dT/dz <0$
in a homogeneous medium, i.e., $d\mu/dz = 0$.  In
panel (e) in Figure~\ref{fig:physics} we consider the situation
where $\tdi \gg \od$. Under this condition, a fluid element that is 
displaced with  $\xi_z >0$ ($\xi_z <0$) is effectively heated (cooled) by conduction
along the magnetic field lines which have been distorted by the displacement 
of the fluid element. This causes the fluid element to expand (contract), and thus
attain a density which is lower (higher) than the surrounding medium. This
leads to the runaway process known as the MTI. The term proportional to the composition
gradient is responsible for the onset of the MTCI when $d\mu/dz >0$ in an isothermal 
environment. This is illustrated in  panel (f) in Figure~\ref{fig:physics}, 
which shows that because of the effective diffusion of ions along distorted 
magnetic field lines, fluid elements that have been displaced upward (downward)
$\xi_z >0$ ($\xi_z <0$) maintain the mean molecular weight corresponding to the equilibrium value 
at their original position. This implies that the fluid element is immersed in a medium 
that is relatively denser (lighter) and it will thus rise (sink).
When both gradients are non-zero their relative magnitudes determine whether the plasma
is buoyantly unstable according to Equation~(\ref{eq:delta_rho_MTCI}). This can also be
seen in the equation of motion
\be
\frac{d^2\xi_z}{dt^2} &=& -N_{T/\mu}^2  \, \xi_z  \,,
\en
which has exponentially growing solutions when $N_{T/\mu}^2<0$.

\subsection{Overstable Modes}

The physics driving overstable modes, which are present when $b_z=1$, 
is more subtle, but some insight can be gained by analyzing directly 
the equations of motion for the corresponding Lagrangian displacements.
Using Equations~(\ref{eq:OSHPBI}) and (\ref{eq:fast_overstable_Bz}) 
together with the equations for the perturbations (\ref{eq:deltav})--(\ref{eq:deltamu})
we find, to first order,
\be
\frac{d^2\xi_z}{dt^2} &=&  N^2_{T\mu}\xi_z -
\frac{N^2+N^2_{T\mu}}{\tci}\frac{d \xi_z}{d t} \,,\,\,\,
\quad(\tci\gg\omega_{\rm dyn}) \,; \qquad \\
\frac{d^2\xi_z}{dt^2} &=& -N^2 \xi_z -\tci 
\frac{N^2+N^2_{T\mu}}{N^2(k_\bot^2/k^2)}\frac{d \xi_z}{d t} \,,
(\omega_{\rm dyn}\gg\tci) \,.\quad\qquad 
\en
The physics behind the first terms on the right-hand side of each 
of these equations is readily recognized.  The term proportional 
to $N^2_{T\mu}$ is responsible for the HPBI if  $N^2_{T\mu}>0$, 
while the term proportional to $N^2$ is responsible for 
Brunt-Vi\"as\"al\"a oscillations, i.e., stable $g$-modes, if $N^2>0$.
In a medium where $N^2_{T\mu}<0$ and $N^2>0$ but $N^2+N^2_{T\mu}<0$, 
the anisotropic heat flow along magnetic field lines tends to monotonically increase
the restoring force acting on the oscillating fluid parcel and leads to overstability.

\section{Astrophysical Implications}
\label{sec:discussion}

Throughout this paper, we have studied the various instabilities
that can be present in the parameter space spanned by $(d\ln \mu/d\ln P,
d\ln T/d\ln P)$ without imposing restrictions on the relative values
of the gradients involved.  We can now frame our results in the context 
provided by observations and theoretical models addressing the 
temperature and composition structure of galaxy clusters.

\subsection{Temperature and Composition Profiles in the ICM
from Models and Observations}

Modern X-ray observatories, such as \emph{Chandra} 
and \emph{XMM-Newton}, have made it possible to obtain the temperature 
profiles of a large number of galaxy clusters, see, e.g., 
\citet{2006ApJ...640..691V, 2008A&A...486..359L}.
This has enabled to show that, quite generically, the gas temperature 
increases with radius, reaching values of 5 -- 10 keV at 100 kpc, and 
decreases toward  the outer cluster regions. 
In several cases the temperature changes by a factor of two or three
on distances comparable to the cluster radius. 

In spite of the small mass ratio between hydrogen and helium, 
because helium is relatively abundant in the primordial gas from which 
clusters form, its sedimentation over the life-time of the
cluster has the potential to produce important gradients
in the mean molecular weight profile. 
Obtaining observational evidence to quantify the helium abundances
that would result form this sedimentation is very hard because
helium is completely ionized at the characteristic temperatures 
of typical galaxy clusters. However, this information is key in order
to derive physical cluster properties, such as gas mass, total mass 
and gas mass fraction (see, e.g., \citealt{2000ApJ...529L...1Q}), and 
even cluster distances \citep{2007arXiv0705.3289M} from X-ray observations.
This has highlighted the need to understand the efficiency
of this process on theoretical and observational grounds
\citep{1977MNRAS.181P...5F, 1981ApJ...248..429A, 1984SvAL...10..137G, 
2000ApJ...529L...1Q, 2003MNRAS.342L...5C, 2004MNRAS.349L..13C, 
2004A&A...420..135T, 2006MNRAS.369L..42E,2009ApJ...693..839P}. 
Some of these estimates predict  an overabundance by up to a factor of a 
few with respect to the primordial value $\mu \simeq 0.6$.


\begin{figure}[]
\begin{center}
  \includegraphics[width=0.475\textwidth,trim=0 10 0 -20]{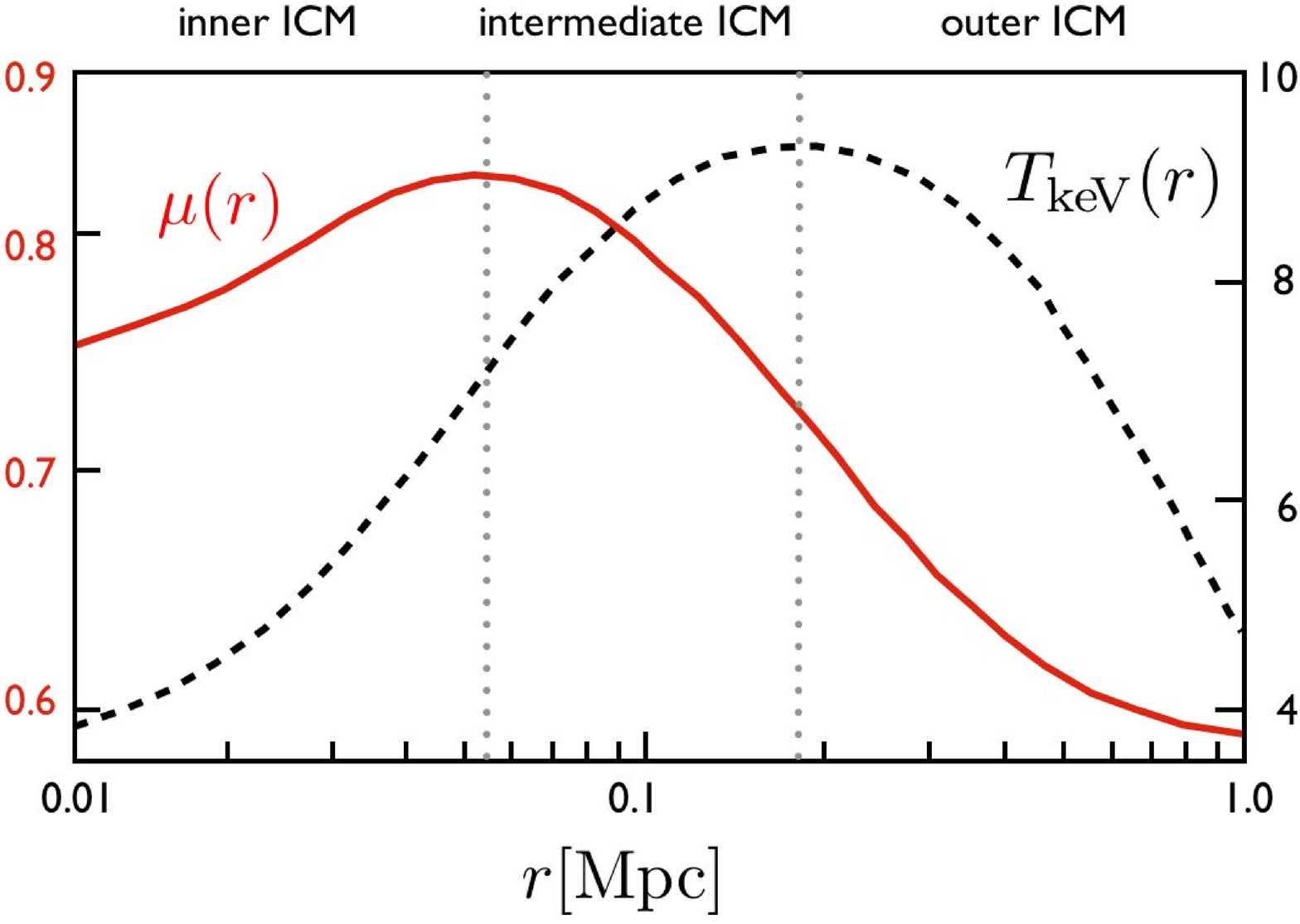}
  \caption{Schematic representation of the mean molecular weight (red
    solid line) and temperature (black dashed line) profiles of a
    representative galaxy cluster as suggested by observations
    \citep{2006ApJ...640..691V} and theoretical models
    \citep{2011A&A...533A...6B}. The regions denoted by ``inner",
    ``intermediate", and ``outer" ICM (delimited by dotted gray lines)
    correspond to three different quarters in the $(d\ln \mu/d\ln P, d\ln T/d\ln P)$ 
    plane (Figures~\ref{fig:fast-conduction}, \ref{fig:slow-conduction}, and panel (a)
    in Figure~\ref{fig:icm-regions}).  The mean molecular weight for a homogeneous 
    cluster with primordial abundance is $\mu \simeq 0.6$.}
  \label{fig:icm-gradients}
\end{center}
\end{figure}

Since the realization that the stability properties of a weakly
magnetized medium depends critically on its thermal structure,
a large number of works have been devoted to study the HBI and the MTI.
In spite of the fact that both instabilities grow on a dynamical timescale,
the onset of the most relevant modes depends explicitly on the local
values of the thermal gradients. Throughout this study, we have 
seen that if the composition of the plasma is not homogeneous,
it is the combination of both the thermal and the mean molecular weight
gradients that decides whether the plasma is stable or not.
In order to understand how important the contributions from each of these
gradients can be, we consider the schematic representation of
the temperature and the mean molecular weight profiles of the 
representative galaxy cluster shown in Figure~\ref{fig:icm-gradients}.
The temperature profile shown resembles the results obtained by 
observations \citep{2006ApJ...640..691V}, whereas
the mean molecular weight profile is akin to the helium sedimentation
models discusses in \citet{2011A&A...533A...6B}, which are based on analytical
models for the physical properties of the ICM introduced in
\citet{2010ApJ...720.1038B}.  
Although very crude, these representative profiles allow us to provide an 
estimate for the values of the logarithmic gradients for the temperature, 
$(\nabla T)/T$, and the mean molecular weight, $(\nabla \mu)/\mu$, which play 
a role in determining the stability of the dilute ICM.
If the peak in the temperature profile occurs at a larger radius than 
the peak in the mean molecular weight profile, as shown in 
Figure~\ref{fig:icm-gradients}, then there are three distinct 
regions defined by the signs of the temperature and the composition gradients. 
Each of these regions of the ICM will have associated different characteristic 
values, which can be estimated according to
\be
\frac{\nabla T}{T} \simeq \frac{1}{L} \frac{\Delta T}{\bar{T}} \,, \qquad
\frac{\nabla \mu}{\mu} \simeq \frac{1}{L} \frac{\Delta \mu}{\bar{\mu}} \,.
\label{eq:gradT_gradmu}
\en
Here, $\Delta T \equiv T_{\rm out} - T_{\rm in}$ stands for the difference 
between two values across the characteristic scale $L \equiv r_{\rm out} - r_{\rm in}$, 
and $\bar{T} \equiv (T_{\rm out} + T_{\rm in})/2$ is the associated  mean value, 
with similar definitions for the mean molecular weight.
Using the information available in Figure~\ref{fig:icm-gradients}, we estimate
these characteristic values for the different ICM regions in Section
\ref{sec:stability_icm}.


\begin{deluxetable*}{lrrrrccc}
\tablecaption{Representative Parameter Values for Different ICM Regions 
\label{table:icm-regions}}
\tablehead{
\colhead{ICM Region}&
\colhead{$\nabla T$}&
\colhead{$\nabla \mu$}&
\colhead{$\beta$}&
\colhead{$K_{n}^{-1}$}&
\colhead{Local Modes in Fluid Model}&
\colhead{Fast Conduction and $\omega_{\rm A} < \od$} &
\colhead{Slow Conduction and $\omega_{\rm A} < \tci$} \\
\colhead{}&
\colhead{}&
\colhead{}&
\colhead{}&
\colhead{}&
\colhead{$K_{n}^{1/2}<\tilde{k}_\parallel<K_{n}^{-1/2}$}&
\colhead{$1/3<\tilde{k}_\parallel<(\beta K_{n})^{1/2}$}&
\colhead{$0.1(\beta K_{n})^{-1/2}<\tilde{k}_\parallel<1/3$}
}
\startdata
Outer-ICM & $<0$ & $<0$ & $10^4$ & $10^{1}$ & $0.3< \tilde{k}_\parallel < 3$ &  $1/3<\tilde{k}_\parallel < 30$ &  $0.003< \tilde{k}_\parallel < 1/3$ \\
Interm.-ICM & $>0$ & $<0$ & $10^3$ & $10^{2}$ & $0.1< \tilde{k}_\parallel < 10$ & $1/3< \tilde{k}_\parallel <  3$  & $0.03< \tilde{k}_\parallel< 1/3$  \\
Inner-ICM & $>0$ & $>0$ & $10^2$ & $10^{3}$ & $0.03< \tilde{k}_\parallel< 30$ & $1/3< \tilde{k}_\parallel< 0.3$  & $0.3< \tilde{k}_\parallel < 1/3$
\enddata
\tablecomments{The various ICM regions are defined in Figure~\ref{fig:icm-gradients}.
For convenience, in this table we have defined the dimensionless parallel wavenumber, 
$\tilde{k}_\parallel = k_\parallel (\lambda_{\rm mfp} H)^{1/2}$.
}
\end{deluxetable*}

\subsection{Applicability of Approximations in the ICM}

Before addressing the stability of the different regions of the ICM,
we comment on two of the approximations that we have made
on the geometry and strength of the magnetic field,  which have allowed us 
to gain physical insight while keeping the problem  tractable.

We have studied in detail two special cases for the orientation 
of the background magnetic field, viz., either parallel, 
$\bb{B} = B \hat{\bb{z}}$, or perpendicular, $\bb{B} = B \hat{\bb{x}}$, 
to the gravitational field $\bb{g}$. These two configurations, which 
have received a lot of attention in the related literature, have the advantage of 
not only simplifying the mathematics involved but also exposing in a clean way the 
physics driving the HBI, the MTI, as well as the generalizations that result from 
including composition gradients. While beyond the scope of this paper, 
accounting for more general geometries is  clearly necessary in order to describe 
more realistic situations.

An important simplification in our study is the assumption
that the magnetic field is so weak that the Alfv{\'e}n frequency is much 
smaller than any other inverse timescale involved. 
It is worth mentioning that this is the regime explored by a number of
numerical studies addressing both fundamental aspects of the MTI and the HBI
\citep{2005ApJ...633..334P, 2007ApJ...664..135P, 2008ApJ...677L...9P, 2012MNRAS.423.1964L},
as well as the implications that these instabilities have for the
long-term evolution of the ICM \citep{2009ApJ...704..211B, 2010ApJ...713.1332R, 2008ApJ...688..905P, 2009ApJ...703...96P, 2010ApJ...712L.194P, 2012MNRAS.422..704P, 2011MNRAS.413.1295M, 
2012MNRAS.419.3319M, 2012ApJ...754..122K}. 
For the modes for which magnetic  tension cannot be neglected, 
the Alfv{\'e}n timescale can become comparable 
or even faster than the dynamical and the conduction timescales and neglecting 
$\omega_{\rm A}$ in the dispersion relation (\ref{eq:fulldisp}) is not a 
good approximation. 

For the modes for which magnetic tension is important, the explicit dependence on 
$\omega_{\rm A}$ might introduce new stability criteria which have not been
captured by our analysis. Furthermore, magnetic tension 
could affect the growth rates of the instabilities. \citet{2011MNRAS.417..602K}
provides a summary of the stabilizing effects provided by 
magnetic tension on HBI- and MTI-unstable modes in a homogeneous medium.
The main physical effect introduced by a non-zero Alfv{\'e}n frequency
is to provide a cut-off for the growth of unstable modes at parallel wavenumbers
such that $k_\parallel v_{\rm A}$ is comparable to the growth rate of the
most unstable modes. For the magnetic field geometries that we analyzed, this must also
be the case even in the presence of a composition stratification. The reason 
for this is that, when either $b_x=1$ or $b_z=1$, all the contributions
introduced by the mean molecular weight gradient appear in the form 
$d\ln T/dz\pm d\ln \mu/dz$. Thus, while
the growth rates and range of unstable modes are affected because of the
changes in the background composition, the effects of non-negligible magnetic
tension on these modes, i.e., the existence of a cut-off parallel wavenumber, 
must be similar to what has been found in the case of a homogeneous medium.
Since the plasma $\beta$ in galaxy clusters varies with radius,  
whether neglecting magnetic tension is a sensitive approximation
for a local stability analysis or not, depends on the conditions 
present in the region of the ICM under consideration. We address this
issue in detail below.


\begin{figure*}[]
\begin{center}
  \includegraphics[width=0.475\textwidth,trim=0 10 0 -20]{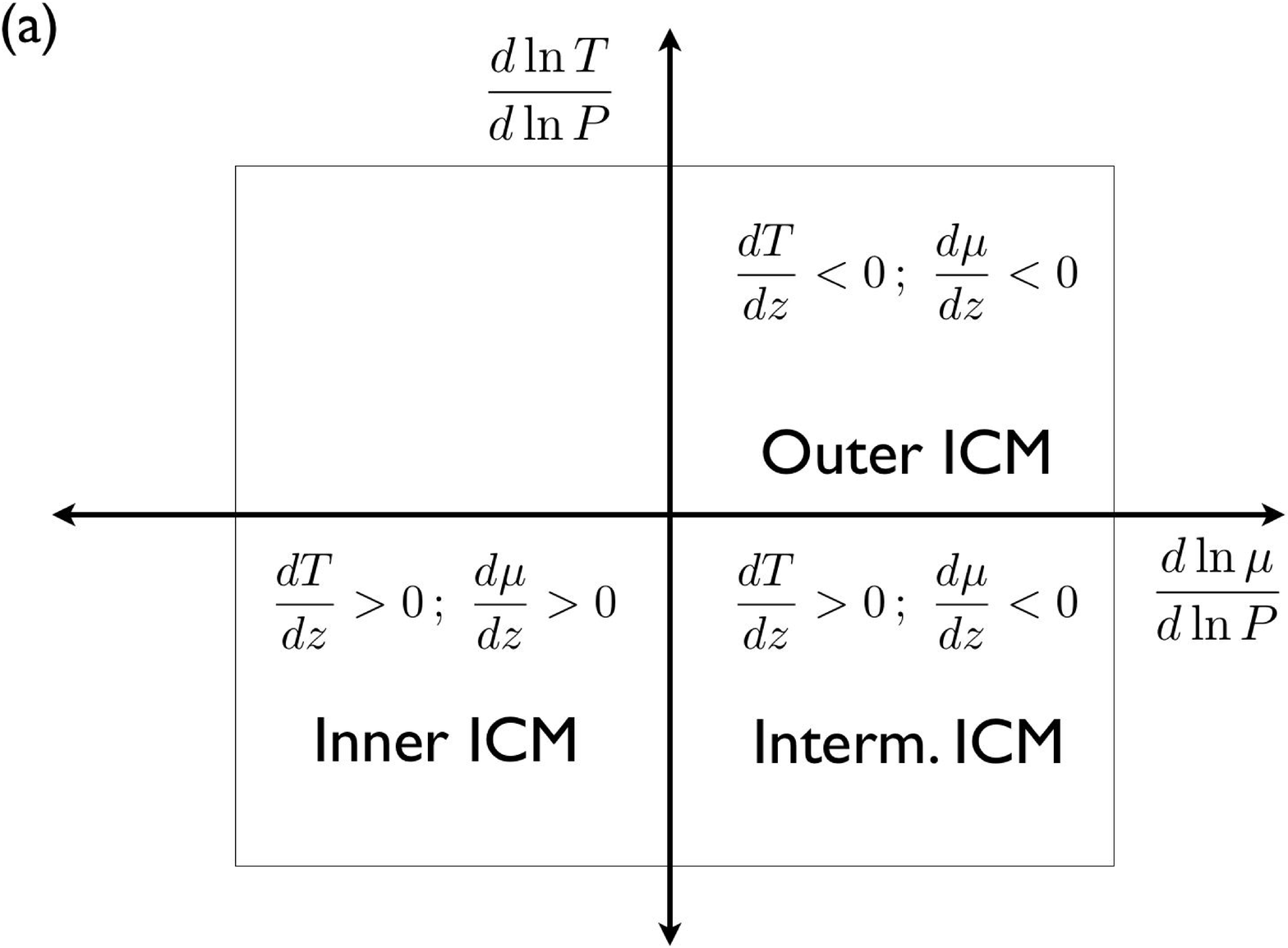}
  \includegraphics[width=0.475\textwidth,trim=0 10 0 -20]{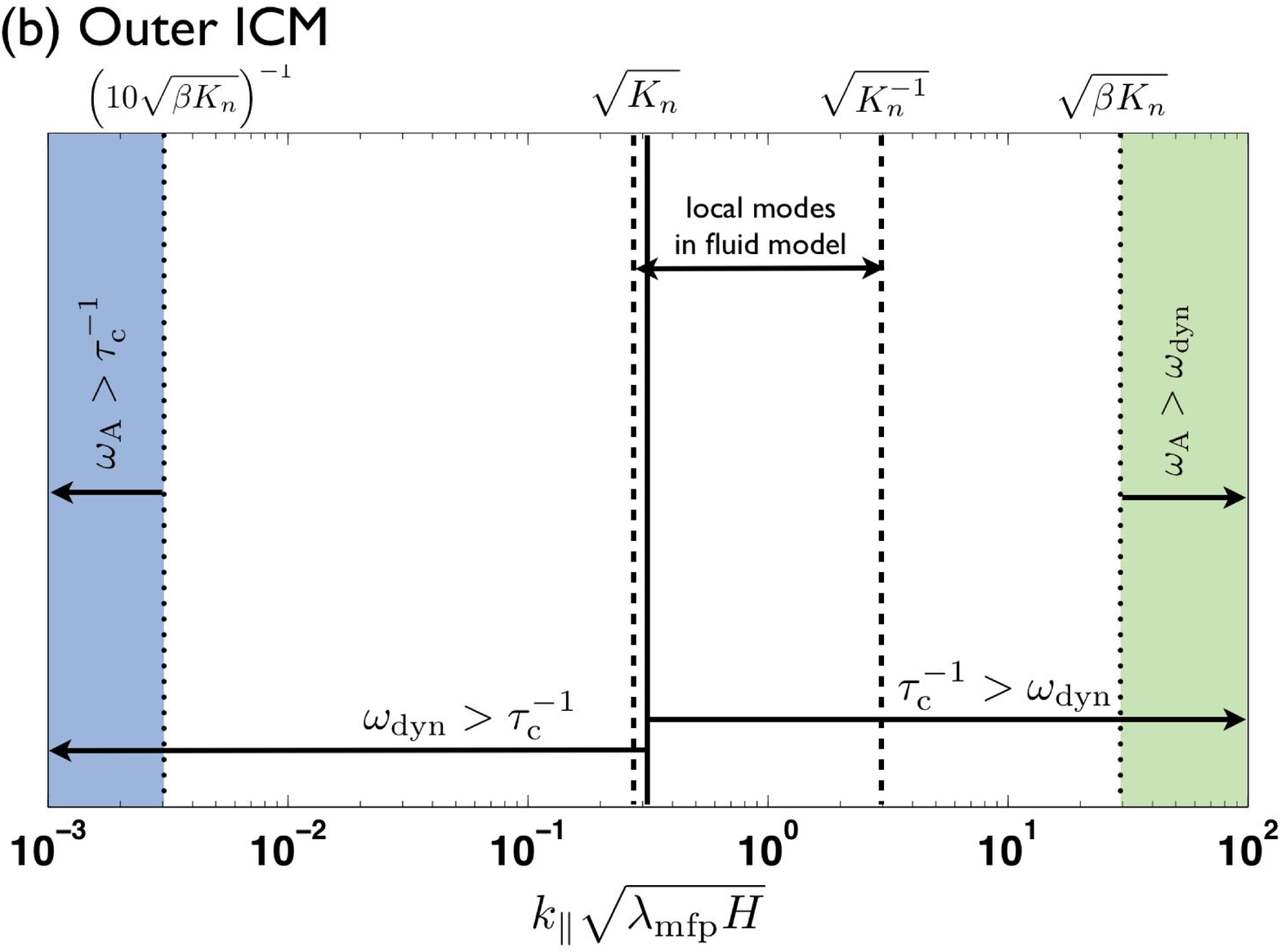}
  \includegraphics[width=0.475\textwidth,trim=0 10 0 -20]{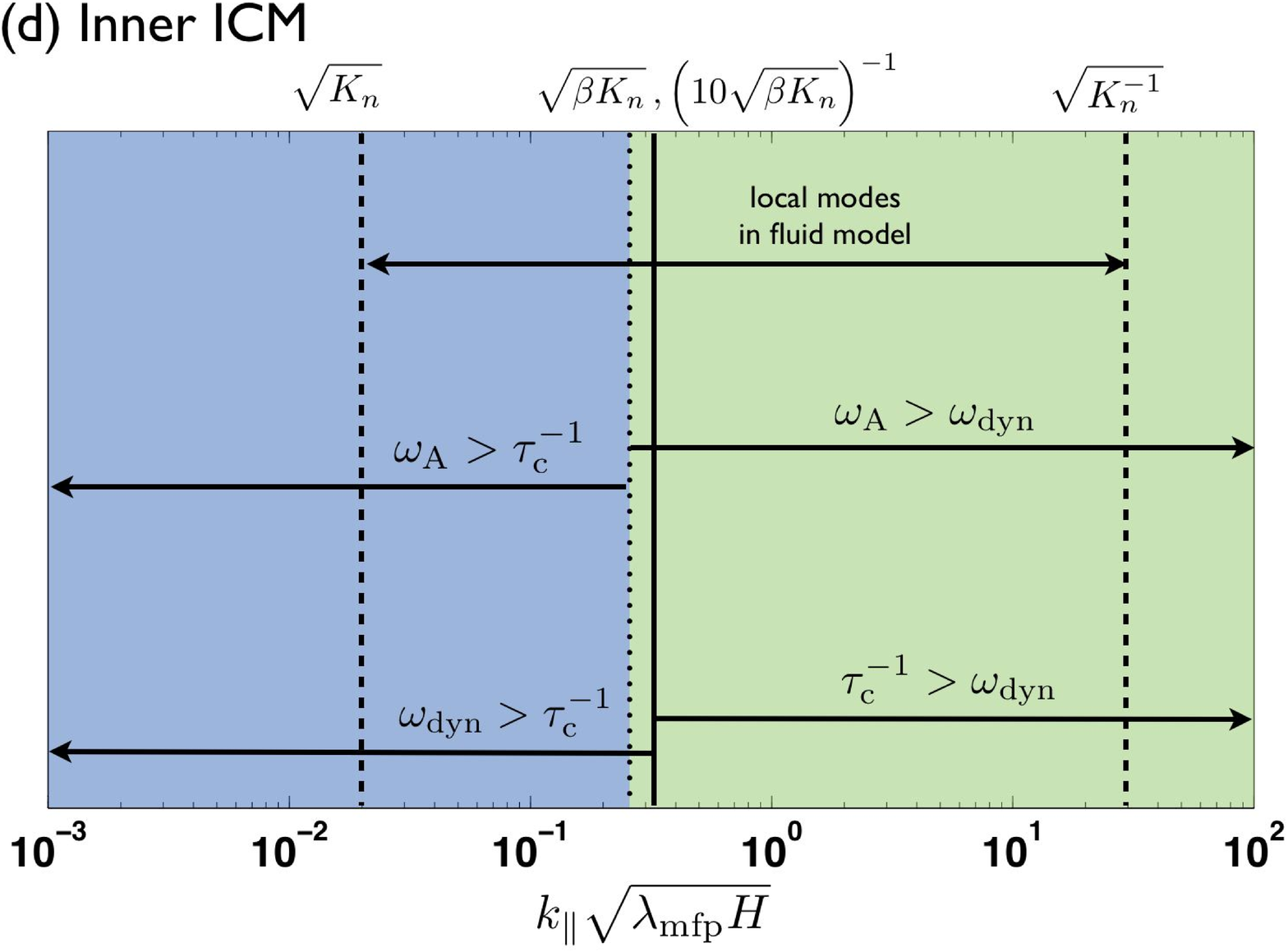}
  \includegraphics[width=0.475\textwidth,trim=0 10 0 -20]{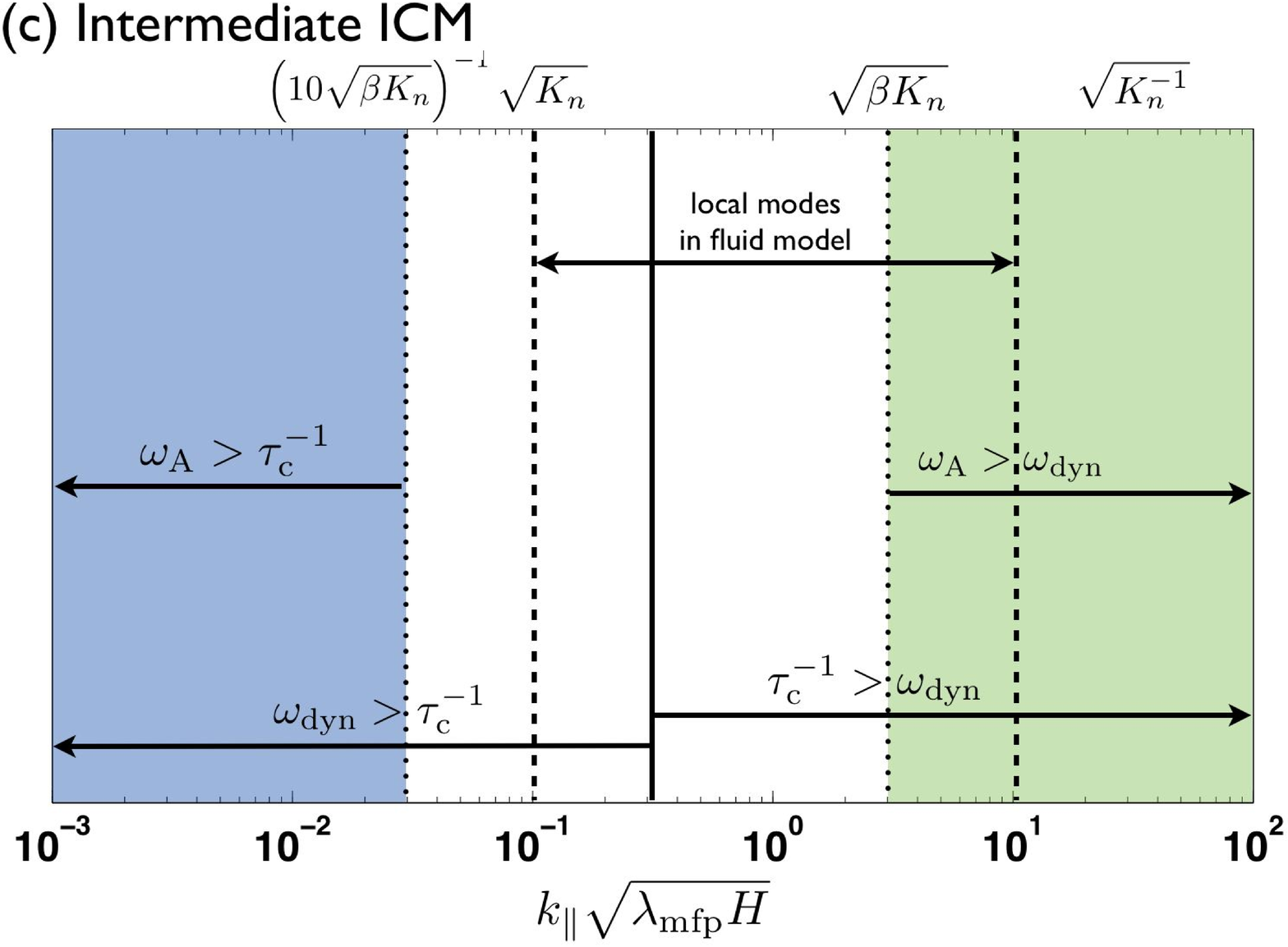}
  \caption{
    Panel (a) shows a schematic representation
    of the correspondence between the regions of a representative
    galaxy cluster with the radial temperature and the mean molecular weight
    profiles as shown in Figures~\ref{fig:icm-gradients} and the plane spanned by 
    $(d\ln T/d\ln P, d\ln \mu/d\ln P)$. 
The various vertical lines in panels (b)--(d) mark some relevant values of the 
dimensionless parallel wavenumber, $k_\parallel (\lambda_{\rm mfp} H)^{1/2}$. 
The region between the dashed lines corresponds to the range of local modes for which the fluid 
approximation is valid, i.e., $K_{\rm n}^{1/2} < k_\parallel (\lambda_{\rm mfp} H)^{1/2} < K_{\rm n}^{-1/2}$.
The solid line, at $k_\parallel (\lambda_{\rm mfp} H)^{1/2}=1/3$, distinguishes between
modes for which conduction timescale is smaller (to the right) or larger (to the left) than the 
dynamical timescale. The dotted lines represent the value of the $k_\parallel (\lambda_{\rm mfp} H)^{1/2}$ 
beyond which magnetic tension cannot be neglected. This corresponds to $k_\parallel (\lambda_{\rm mfp} H)^{1/2} > 
(\beta K_n)^{1/2}$ (green) for fast conduction modes, and $k_\parallel (\lambda_{\rm mfp} H)^{1/2} < 
(\beta K_n)^{-1/2}/10$ (blue) for slow conduction modes.
All the numerical values associated with the various length scales and timescales have been drawn 
from Table~\ref{table:icm-regions}, which provides estimates for the $\beta$ plasma and the 
Knudsen number representative of the different regions of the ICM.
     }
  \label{fig:icm-regions}
\end{center}
\end{figure*}

\subsection{Stability of ICM regions}
\label{sec:stability_icm}

The analysis of Figures~\ref{fig:fast-conduction} and
\ref{fig:slow-conduction} allows us to understand the implications
that a mean molecular weight gradient can have for the various regions
of a representative galaxy cluster as depicted in
Figure~\ref{fig:icm-gradients}.  These regions correspond to
different quadrants in the $(d\ln \mu/d\ln P, d\ln T/d\ln P)$ plane as
shown in panel (a) of Figure~\ref{fig:icm-regions}, which we 
have denoted as inner, intermediate, and outer ICM.
In what follows we will assume, as suggested by observations of galaxy clusters, 
that the ICM is buoyantly stable according to the classical stability criterion
$N^2>0$ \citep{2005A&A...433..101P, 2009ApJS..182...12C}. 

\subsubsection{Outer ICM}

In this region the temperature and the mean molecular weight gradients are both negative.
Because the criteria for stability are different whether conduction is fast or slow
compared to the dynamical timescale, we consider these two cases separately.

\emph{Fast conduction}.
If the magnetic field lines are parallel to the gravitational field, 
i.e., $b_z=  1$, gravity modes can become overstable if $d\ln T/d\ln P >  
(\gamma-1)/(2\gamma)$, panel (a) in Figure~\ref{fig:fast-conduction}. 
A gradient in the mean molecular weight alone is unable to stabilize these modes
and can drive unstable modes driven by diffusion if $|\nabla \mu|/\mu > |\nabla T|/T$,
panels (b) and (c) in Figure~\ref{fig:fast-conduction}.
For magnetic field configurations that are perpendicular to the gravitational field, 
i.e., $b_x=  1$, this region is unstable to the MTCI provided that 
$|\nabla \mu|/\mu < |\nabla T|/T$, panel (d) in Figure~\ref{fig:fast-conduction}. 
This means that the outskirts of galaxy clusters that would 
be considered prone to the MTI (if they were homogeneous) would remain stable 
if the gradient in mean molecular weight is steep enough. This is not the case 
for the particular profiles shown in Figure~\ref{fig:icm-gradients}
but this does not imply that this is not the case in general.

\emph{Slow conduction}.
In the absence of ion-diffusion, gravity modes can become overstable
if bz = 1. These modes cannot be stabilized by means of a gradient in the
mean molecular weight alone, panel (a) in Figure~\ref{fig:slow-conduction}. 
Furthermore, when ion-diffusion is efficient, it can drive 
unstable modes if $|\nabla \mu|/\mu > |\nabla T|/T$, panel (c) in 
Figure~\ref{fig:slow-conduction}. For $b_x=  1$ there can be unstable modes 
driven by conduction if  $|\nabla \mu|/\mu < |\nabla T|/T$,
panel (d) in Figure~\ref{fig:slow-conduction}.

We can provide a crude estimate of the impact that a composition gradient
would have on the growth rates of the various instabilities discussed by estimating 
the temperature and the composition gradients shown in Figure~\ref{fig:icm-gradients}. 
For the inner ICM region, the characteristic scale is $L \simeq 0.8$ Mpc, while
$\Delta T \simeq 4$ keV, $\bar{T} \simeq 7$ keV, $\Delta \mu \simeq 0.1$, and 
$\bar{\mu} \simeq 0.65$. Thus according to Equations (\ref{eq:gradT_gradmu}),
the characteristic logarithmic gradients in this inner region are given by
\be
\left.\frac{\nabla T}{T} \right|_{\rm outer}    \simeq -0.7 \,\textrm{Mpc}^{-1}\,, \,\,
\left.\frac{\nabla \mu}{\mu} \right|_{\rm outer}\simeq -0.2 \, \textrm{Mpc}^{-1} \,.\qquad
\en
These order-of-magnitude estimates, based on the representative values drawn from
Figure~\ref{fig:icm-gradients}, show that the instabilities with growth rates
$\sigma^2 \propto -\ln (T/\mu)$, such as the generalization of the MTI, 
Equations~(\ref{eq:sigma_sq-MTCI}) and (\ref{eq:sigma_sq-DMTCI}), will be 
15\% slower with respect to the homogeneous case.

Regarding the validity of our assumptions of weak magnetic fields, 
panel (b) of Figure~\ref{fig:icm-regions} shows that magnetic tension is unimportant
for the range of modes of interest, i.e., 
$K_{\rm n}^{1/2} < k_\parallel (\lambda_{\rm mfp} H)^{1/2} < K_{\rm n}^{-1/2}$. Thus, our
approximation of setting $\omega_{\rm A} \simeq  0$ is fully justified in this region.

\subsubsection{Intermediate ICM}

In this region the temperature and the mean molecular weight gradients have different 
signs, \footnote{Note that if the peak in $\mu$ were to occur beyond the radius at which 
the temperature is maximum, then this region would be characterized by $dT/dz <0$ and $d\mu/dz>0$, 
which corresponds to the opposite quadrant in panel (a) of Figure~\ref{fig:icm-regions}.} with 
$\nabla \mu<0$ and  $\nabla T>0$ according to the profiles shown in Figure~\ref{fig:icm-gradients}.

When $b_z=  1$, this region is unstable due to the HPBI, which grows on the dynamical
timescale if $|\nabla \mu|/\mu < |\nabla T|/T$. This instability can be prevented if the mean 
molecular weight is steep enough, panel (a) in Figure~\ref{fig:fast-conduction}. 
When ion-diffusion is efficient this region is unstable due to the D-HPBI,
panel (c) in Figure~\ref{fig:fast-conduction}. 
This magnetic field configuration is also prone to unstable modes for which conduction is slow. 
In this case there are unstable modes driven by conduction if $D=0$ and $|\nabla \mu|/\mu < 
|\nabla T|/T$, while there are unstable modes driven by ion-diffusion if $D\ne0$ and 
$|\nabla \mu|/\mu > |\nabla T|/T$, panels (b) and (c) in Figure~\ref{fig:fast-conduction}, 
respectively. If $b_x=  1$, this region is stable whether conduction is fast or slow compared 
to the dynamical timescale, panels (d) in Figure~\ref{fig:fast-conduction} and (e) in 
Figure~\ref{fig:slow-conduction}.

For this intermediate ICM region, the inspection of Figure~\ref{fig:icm-gradients} 
provides $L \simeq 0.15$ Mpc, $\Delta T \simeq 2$ keV, $\bar{T} \simeq 8$ keV,
$\Delta \mu \simeq 0.1$, and $\bar{\mu} \simeq 0.75$. Thus the characteristic 
logarithmic gradients are given by
\be
\left.\frac{\nabla T}{T} \right|_{\rm interm.} \simeq 1.6 \,\textrm{Mpc}^{-1}\,, \,\,
\left.\frac{\nabla \mu}{\mu} \right|_{\rm interm.}\simeq -1 \, \textrm{Mpc}^{-1} \,.\qquad
\en
Therefore, the instabilities with growth rates for which
$\sigma^2 \propto \ln (T\mu)$, such as the generalization of the
HBI in the absence of ion-diffusion, Equation~(\ref{eq:sigma_sq-HPBI}),  
will grow of the order of 40\% slower with respect to the homogeneous case.
On the other hand, the instabilities  with growth rates $\sigma \propto \ln (T/\mu)$, 
such as the generalization of the HBI with ion-diffusion, Equation~(\ref{eq:sigma_sq-DHPBI}), 
will be 30\% faster.

As shown in panel (c) in Figure~\ref{fig:icm-regions}, 
magnetic tension is important for some of the modes for which conduction is fast
but not for the modes for which conduction is slow.

\subsubsection{Inner ICM}

In this region the temperature and the mean molecular weight gradients are both positive.
We consider again the limits in which conduction is fast or slow separately.

\emph{Fast conduction}.
For $b_z=  1$, this region is unstable to the HPBI regardless of whether the mean 
molecular weight gradient is smaller or larger than the temperature gradient, panel (a) in 
Figure~\ref{fig:fast-conduction}. Furthermore, if ion-diffusion is efficient it can also drive
unstable modes, panels (b) and (c) in  Figure~\ref{fig:fast-conduction}.
In a homogeneous medium with $b_x=  1$, this inner region is stable against the MTI. 
However, there can be unstable MTCI-modes if $|\nabla \mu|/\mu > |\nabla T|/T$, panel (d) in 
Figure~\ref{fig:fast-conduction}. 
In the homogeneous case when $\nabla T>0$, the HBI tends to re-orient the magnetic field 
in the radial direction, which results in a field configuration which is stable against the MTI, 
i.e., $b_x \simeq 1$. When ion-diffusion is not efficient, this core insulation could be 
alleviated by the MTCI if the mean molecular weight gradient is steep enough, i.e., 
$(\nabla \mu)/\mu > (\nabla T)/T$.

\emph{Slow conduction}.
This region can be subject to instabilities driven by both heat conduction and ion-diffusion.
When $b_z=  1$ and ion-diffusion is inefficient, there are unstable modes driven by heat conduction
regardless of the relative magnitude of the temperature and the mean molecular weight profiles,
panel (b) in Figure~\ref{fig:slow-conduction}, whereas there are unstable modes driven by 
diffusion if $|\nabla \mu|/\mu > |\nabla T|/T$, panel (c) in Figure~\ref{fig:slow-conduction}.
In the case with $b_x=  1$, there can be unstable modes driven by heat conduction if 
$|\nabla \mu|/\mu > |\nabla T|/T$, panel (e) in Figure~\ref{fig:slow-conduction}. 

According to Figure~\ref{fig:icm-gradients}, the inner ICM region is characterized by
$L \simeq 0.05$ Mpc, $\Delta T \simeq 3$ keV, $\bar{T} \simeq 5.5$ keV,
$\Delta \mu \simeq 0.1$, and $\bar{\mu} \simeq 0.8$, and thus
\be
\left.\frac{\nabla T}{T} \right|_{\rm inner} \simeq 10 \,\textrm{Mpc}^{-1}\,, \,\,
\left.\frac{\nabla \mu}{\mu} \right|_{\rm inner}\simeq 2.5 \, \textrm{Mpc}^{-1} \,.\qquad
\en
Therefore, the instabilities with growth rates for which
$\sigma^2 \propto \ln (T\mu)$ will grow of the order of 10\% faster with 
respect to the homogeneous case, while the instabilities 
with growth rates $\sigma^2 \propto \ln (T/\mu)$ will be 15\% slower.

The analysis of Figure~\ref{fig:icm-regions} shows that the range of modes for which it is 
sensible to carry out a local mode analysis within the fluid model embodied in 
Equations~(\ref{eq:rho})--(\ref{eq:c}) increases as the inverse Knudsen number increases 
toward  smaller radii. However, due to the increase in the strength of the 
background magnetic field, the range of modes for which it is sensible to neglect magnetic 
tension, decreases. Because the relatively low values of $\beta$ in the inner core region, 
magnetic tension is important for all the modes of interest. Therefore, our conclusions
for this region should be considered with caution. 


\begin{figure*}[]
\begin{center}
  \includegraphics[width=\textwidth,trim=0 10 0 -20]{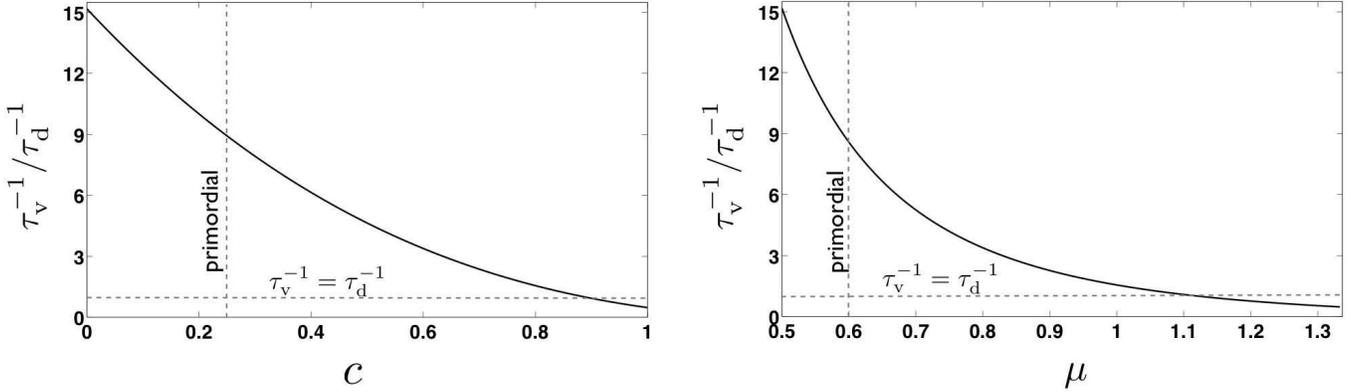}
  \caption{Ratio of the inverse timescales associated with viscous and diffusion processes 
  for a binary mixture of H and He as a function of the composition $c$ (left) and mean 
  molecular weight $\mu$ (right). The vertical dashed line indicates the values for which 
  the mixture has primordial composition, i.e., $c \simeq  0.25$, which corresponds to 
  $\mu \simeq  0.6$.}
  \label{fig:tvi-tdi}
\end{center}
\end{figure*}

\subsubsection{Summary and Outlook}

During the last few years, there has been substantial numerical work 
for understanding the long-term evolution of the MTI and the HBI and their implications
for the gas dynamics in the ICM permeating galaxy clusters
\citep{2009ApJ...704..211B, 2008ApJ...688..905P, 2009ApJ...703...96P, 
2010ApJ...712L.194P, 2012MNRAS.422..704P, 2011MNRAS.413.1295M, 2012MNRAS.419.3319M, 
2012ApJ...754..122K}.
All of this work has been done under the assumption that the ICM is homogeneous
and thus the temperature gradient provides the only source of energy to feed
instabilities. Even though it is hard to quantify concentration gradients
from observations, some heavy element sedimentation is expected
\citep{2001ApJ...562L.129N,  2003MNRAS.342L...5C, 2004MNRAS.349L..13C, 2006MNRAS.369L..42E}. 
Indeed, current theoretical models suggest that helium sedimentation can significantly 
alter the composition profile throughout the cluster and give rise to mean molecular weight 
gradients which are comparable in magnitude to the temperature gradients, with 
$|\nabla T|/T \simeq |\nabla \mu|/\mu$ (see, e.g., \citealt{2011A&A...533A...6B}, 
and Figure~\ref{fig:icm-gradients}). 

This work discusses for the first time the effects that composition gradients
can have for the stability of a weakly collisional magnetized medium 
which is stratified in both temperature and composition. We have found that, 
depending on the wavelength of the modes under consideration, the plasma can be 
subject to a wide variety of unstable modes. These include:
\begin{itemize}
\item  the generalization of  the MTI \citep{2001ApJ...562..909B}:
\be
\sigma^2& \approx& -g\f{d\ln (T/\mu)}{dz}\f{k_x^2+k_y^2}{k^2}\,,
\en
\end{itemize}
and the generalization of the HBI \citep{2008ApJ...673..758Q} in
\begin{itemize}
\item the slow ion-diffusion limit:
\be
\sigma^2 &\approx& g\frac{d\ln (T\mu)}{dz}\frac{k_\bot^2}{k^2}\,,
\en
\item the fast ion-diffusion limit:
\be 
\sigma \approx \frac{g}{\tvi} \f{d\ln (T/\mu)}{dz} \,.
\en 
\end{itemize}
We have also found the generalization of the overstable gravity modes 
discussed in \citet{2010ApJ...720L..97B}, see Equations~(\ref{eq:OSHPBI}) 
and (\ref{eq:fast_overstable_Bz}), 
as well as other new modes  which are driven by conduction and diffusion. 

This study constitutes a first step toward the long-sought goal of
understanding in a self-consistent way the effects of magnetic turbulence on
the diffusion of heavy elements and its consequences for the observational 
signatures and long-term evolution of galaxy clusters. This will only be 
possible through numerical studies involving realistic models for the microphysics
of weakly collisional, multi-component plasmas. Addressing this problem will 
require to sort out several details, including how to properly handle plasma 
micro-instabilities, many of which are still the subject of active research in 
homogeneous settings.

\acknowledgements

We thank Matthew Kunz, Henrik Latter, Aldo Serenelli, and Shantanu Mukherjee for
useful discussions. We are grateful to the anonymous referee for a 
detailed and thoughtful report that helped us improve 
the final version of this manuscript significantly.  
M.E.P is grateful to the Knud H{\o}jgaard Foundation and the Villum 
Foundation for their generous support.  S.C. acknowledges support from 
the Danish Research Council through  FNU Grant No. 505100-50 - 30,168.

\appendix

\section{Ion Diffusion in a Binary Mixture}
\label{app:tdi-vs-tvi}

The ratio between the timescales associated with viscous and diffusion processes is 
$\tvi/\tdi = 3\nu_\parallel/D$. Here, the coefficient $\nu_\parallel= v_{\rm th}^2/(2\nu_{ii}^{\rm eff})$ 
denotes the kinematic Braginskii viscosity associated with a binary mixture of ions, 
where $\nu_{ii}^{\rm eff}$ is an effective ion-ion collision rate which can be estimated as follows.
The  Braginskii viscosity for a single species of ions is $\eta_0 = \rho v_{\rm th}^2/(2\nu_{ii})$,
where $\nu_{ii}$ is the ion-ion collision frequency 
\begin{eqnarray}
\nu_{ii}=\frac{4\sqrt{\pi}}{3} \frac{n_i q_i^4}{m_i^{1/2}(k_BT)^{3/2}}\ln\Lambda_{ii}\,.
\end{eqnarray}
Here $n_i$, $m_i$ and $q_i$ are the number density, the mass, and the charge of the ion 
respectively; and $\ln \Lambda$ refers to the corresponding Coulomb logarithm. In a binary mixture, 
the effective viscosity coefficient (ignoring the contribution from electrons) is given by
\begin{eqnarray}
\eta_0\simeq  \frac{n_{i_1}k_BT}{\nu_{i_1{i_1}}+\nu_{i_1{i_2}}}+\frac{n_{i_2}k_BT}{\nu_{i_2{i_1}}+\nu_{i_2{i_2}}}\,,\label{eq:eta0A}
\end{eqnarray}
where the collision frequency between ions of species $i$ and $j$ is
\begin{eqnarray}
\nu_{ij}=\frac{4\sqrt{2\pi}}{3}\left[\frac{\sqrt{m_{ij}} 
q_i^2q_j^2 n_j}{m_i(k_BT)^{3/2}}\right]\ln\Lambda_{ij} \,,
\label{eq:nupA}
\end{eqnarray}
and $m_{ij}\equiv m_im_j/(m_i+m_j)$ is the reduced ion mass. We can thus define
the effective ion-ion collision frequency as
\be
\nu_{ii}^{\rm eff}\equiv \frac{\rho v_{\rm th}^2}{2\eta_0}\,.
\en
The coefficient governing the diffusion of species 2 into species 1 
(e.g., helium into hydrogen) is given by \citep{1990ApJ...360..267B}
\begin{eqnarray}
D=&&\frac{3}{4\sqrt{2\pi}} \frac{m_2(k_BT)^{5/2}}{\sqrt{m_{12}}q_1^2q_2^2 \rho\ln\Lambda_{12}} \left[ \frac{4-c}{(2-c)(8-5c)}\right]\,; \nonumber \\ && (0<c<1)\, .
\end{eqnarray}
Therefore the ratio $\tvi/\tdi = 3\nu_\parallel/D$ is given by
\begin{eqnarray}
\frac{\tvi}{\tdi} \simeq && 3\sqrt{\frac{2}{5}}\left[\frac{(2-c)(8-5c)}{4-c}\right] \nonumber \\ 
&& \times
\left[\frac{2}{1+\sqrt{\frac{8}{5}}\left(\frac{c}{1-c}\right)}+\frac{0.25}{1+\sqrt{\frac{2}{5}}\left(\frac{1-c}{c}\right)}\right] \,, \,\,\,\,
\end{eqnarray}
which is shown in Figure~\ref{fig:tvi-tdi} as a function of the concentration $c$ (left panel)
and as a function of the mean molecular weight $\mu$ (right panel). 
In the inner regions of the ICM, where $c\simeq 0.6$ or $\mu \simeq 0.8$ (see Figure~\ref{fig:icm-gradients}),
viscous and diffusion processes take place on comparable timescales, i.e., $\tvi/\tdi \simeq 3$, while 
$\tvi/\tdi \simeq 9$ for a primordial mixture of helium and hydrogen, i.e., $c\simeq  0.25$ or $\mu\simeq 0.6$.
It should be kept in mind that in carrying out this calculation we have assumed that 
all the ratios between the Coulomb logarithms are of order unity and
the expressions for the transport coefficients along the magnetic field lines are 
identical to the ones that are valid in the absence of the magnetic field.


\end{document}